\documentclass{aastex62}

\graphicspath{{./}{figures/}}
\usepackage{amsmath}

\received{January 1, 2018}
\revised{January 7, 2018}
\accepted{\today}
\submitjournal{ApJ}

\shorttitle{CI Tau's Magnetic Field}
\shortauthors{Sokal et al.}

\begin{document}

\title{The Mean Magnetic Field Strength of CI Tau}

\correspondingauthor{Kimberly R. Sokal}
\email{ksokal@utexas.edu}

\author{Kimberly R. Sokal}
\affiliation{Department of Astronomy, The University of Texas at Austin, Austin, TX, 78712, USA}

\author{Christopher M. Johns-Krull}
\affiliation{Department of Physics and Astronomy, Rice University, 6100 Main Street, MS-108, Houston, TX 77005, USA}

\author{Gregory N. Mace}
\affiliation{Department of Astronomy, The University of Texas at Austin, Austin, TX, 78712, USA}

\author{Larissa Nofi}
\affiliation{Lowell Observatory, 1400 W. Mars Hill Rd., Flagstaff, AZ 86001, USA}

\author{L. Prato}
\affiliation{Lowell Observatory, 1400 W. Mars Hill Rd., Flagstaff, AZ 86001, USA}

\author{Jae-Joon Lee}
\affiliation{Korea Astronomy and Space Science Institute, 776 Daedeokdae-ro, Yuseong-gu, Daejeon 34055, Korea}

\author{Daniel T. Jaffe}
\affiliation{Department of Astronomy, The University of Texas at Austin, Austin, TX, 78712, USA}



\begin{abstract}

We present a blind comparison of two methods to measure the mean surface magnetic field strength of the classical T Tauri star CI Tau based on Zeeman broadening of sensitive spectral lines.  Our approach takes advantage of the greater Zeeman broadening at near-infrared compared to optical wavelengths.  We analyze a high signal-to-noise,  high spectral resolution spectrum from 1.5--2.5$\mu$m observed with IGRINS (Immersion GRating INfrared Spectrometer) on the Discovery Channel Telescope. Both stellar parameterization with MoogStokes (which assumes an uniform magnetic field) and modeling with SYNTHMAG (which includes a distribution of magnetic field strengths) yield consistent measurements for the mean magnetic field strength of CI Tau is B of  $\sim$ 2.2 kG. This value is typical compared with measurements for other young T Tauri stars and provides an important contribution to the existing sample given it is the only known developed planetary system hosted by a young classical T Tauri star. Moreover, we potentially identify an interesting and suggestive trend when plotting the effective temperature and the mean magnetic field strength of T Tauri stars. While a larger sample is needed for confirmation, this trend appears only for a subset of the sample, which may have implications regarding the magnetic field generation.

\end{abstract}

\keywords{ infrared: stars --- stars: fundamental parameters --- stars: individual (CI Tau) --- stars: pre-main sequence -- stars: magnetic fields}

\section{Introduction} \label{sec:intro}

Strong stellar magnetic fields play a fundamental role in the
pre-main sequence (PMS) and early main sequence evolution of late type stars.  It is now
well established that the interaction of the newly formed star, also known as
a T Tauri star (TTS), with its disk is strongly regulated by the stellar 
magnetic field.  This interaction is described principally by the 
magnetospheric accretion paradigm \citep{Bou07}.  In 
magnetospheric accretion, the large scale component of the stellar field 
truncates the accretion disk at or near the co-rotation radius, redirecting 
the path of accreting disk material so that it flows along the stellar magnetic
field lines to the surface of the star.  It is usually assumed that the 
footprints of the stellar magnetic field, which take part in the accretion process, are 
anchored at high latitude so that accretion occurs near the stellar poles.  
When the accreting material impacts the stellar surface, it experiences a
strong shock, heating up to $\sim 10^6$ K \citep[][]{Cal98}.  In addition
to accretion of disk material, disk bearing young stars also experience
strong outflows \citep[e.g.][]{Har95, Edw06} that the stellar magnetic field
likely plays an important role in launching \citep[e.g.][]{Shu94, Rom09,
Zan13}.  Emission from the gas taking part in these accretion and outflows 
that are mediated by the stellar magnetic field produce most of the observational
signatures that define classical TTSs (CTTSs).

In addition to mediating the outflows and accretion of disk material onto young stars,
the stellar magnetic field plays a significant role in the rotational
evolution of solar type stars.  As a young star contracts during its PMS
evolution, conservation of angular momentum would result in the rotation
rate of the star increasing.  However, during the CTTS phase, the rotation of
the young star appears to become locked to that of the disk at or near 
the truncation radius \citep[e.g.][]{Edw93, Joh02, Reb06, Cie07, Cau12}.  The
value of the truncation radius depends on the strength and geometry of the 
stellar magnetic field and on the disk accretion rate as well as on the 
stellar mass through its influence on the velocity of orbiting material
\citep[][]{Els77, Gho79, Har98, Bes08}.  While some authors question
whether the action of the stellar magnetic field is sufficient to 
supply the needed torque and enforce disk locking \citep[][]{Uzd02, Mat05,
Mat10, Aar13}, and other studies do not always find a clear observational
signature of disk locking \citep[][]{Reb01, Sta01}, current models of PMS 
angular momentum evolution require disk locking or some similar process in 
order to reproduce the observed distribution of rotation periods in young 
clusters \citep[e.g.][]{Kri97, Bou97, Bar01, Tin02, Irw08, Gal13, Gal15}.
In order to match the range of observed rotation periods in young clusters,
most studies find that disk locking must act over a range of times in
the early evolution of TTSs, consistent with the observed falloff of
disk fraction in young star forming regions \citep[][]{Hai01, Her07, Wya08}.
Once the disks vanish and the young star has contracted to the main sequence,
the stellar field remains important for rotational evolution through the 
action of a magnetized stellar wind which spins the star down over the course
of a few Gyr \citep[e.g.][]{Web67, Sku72, Mat15}. 

In addition to their importance in the rotational evolution of newly formed
stars, strong stellar fields are also critical for our understanding of stellar ages and
have implications for planetary systems.  
Planets form in the disks around young stars
\citep[][]{Joh14, Ray14, Cha14}.  Therefore, the timescale
for planet formation, planet migration, and other processes in the 
protoplanetary disk that help determine the final architecture of planetary
systems is set by the lifetime of the disk.  The disk lifetimes are 
found by measuring the age of newly formed stars \citep[e.g.][]{Her07},
and uncertainties in stellar ages translate directly into uncertainties in
disk lifetimes. A number of studies have now established  that ubiquitous,
strong stellar magnetic fields can measurably alter the structure of low
mass stars \citep[][]{Mul01, Cha07, Mac09, Tor10}, including TTSs \citep[][]{Fei13,
Fei14, Fei16}.  The resulting change in stellar evolution can produce a 
factor of two discrepancy in the age of young stars at the few million year 
timeframe which could possibly explain the difference in age found for some
clusters \citep[][]{Pec12} when the age is determined from the intermediate 
to high mass stars compared to the low mass stars \citep[][]{Fei16}.  As
a result, understanding the magnetic field properties of young stars is
critical for establishing timescales involved in planet formation.
A related effect comes from the influence that magnetic fields can have on
the ability to properly place young stars in an HR diagram and determine
their ages by comparing to model PMS evolutionary tracks.  It is often
desirable to use spectroscopic techniques to measure the effective 
temperature and/or gravity to aid in placing stars in the HR diagram,
and failing to account for strong stellar magnetic fields can still result
in significant systematic offsets on where stars appear in the HR
diagram \citep[e.g.][]{Dop03, Sok18}.  This can be particularly 
important when using near-infrared (NIR) spectra, advantageous for cool and
embedded sources, because the wavelength dependence of the Zeeman
effect makes NIR atomic lines particularly sensitive to, and therefore
strongly affected by, magnetic fields \citep[e.g.][]{Saa85, Joh99, Joh07}.

Furthermore, magnetic fields may have important effects on the planet formation
process itself, for example the observed pileup of planets on very
close orbits and the likelihood of planets
being habitable.  Hot Jupiters, roughly Jupiter mass
planets in very close orbits around their host stars, have a peak in their
distribution at $\sim 0.04$ au \citep[e.g.][]{Bar14, Hel18}.  It is now well
accepted that these planets must form significantly further out in
the disk and migrate in through some mechanism \citep[][]{Lin86, Pap07,
Tri10, Nao11}. The stellar magnetosphere can cause inner disk truncation, which then halts inward migration and leads to a pileup of hot Jupiters
\citep[e.g.][]{Lin96, Cha10,Ida10, Pla13, Bar14}.
High energy radiation
resulting from the stellar magnetic activity also potentially plays an
important role in disk ionization structure and chemistry
\citep[e.g.][]{Gla97, Gla04, Dul10, Ada14}, which influences the environment
where planets form and potentially migrate.  Once planets do form,
the stellar magnetic field likely plays a significant role in the potential 
habitability of worlds around other stars, again through the impact of
high energy radiation resulting from the magnetic activity which is then
incident on planetary atmospheres \citep[e.g.][]{Kal10, Seg10, Til17}.  As
a result, it is important to know the magnetic properties of young stars
that are in the process of forming, or have recently formed, planets.

Planetary mass companions in close orbits around their host stars have so far
been identified through radial velocity (RV) and transit searches.    
Both RV and transit search studies of young stars are significantly
impacted by astrophysical noise resulting from the extreme stellar and
accretion activity of young stars \citep[][]{Pau06, Des07, Hue08, Pra08, 
Mah11}. Hampered by this noise, some claimed detections have later been found to be the
results of stellar activity.  Nevertheless, there have been a number of RV and transit surveys
for planets around young low mass stars 
\citep[][]{Set07, Set08, Her10, van11, Cro12, Bai12, Ngu12, Lag13,
Gag16}.   To date, only a handful of high quality planet candidates in close orbits
around young ($\leq 10$ Myr) parent stars have been announced 
\citep[][]{van12, jk16, Don16, Dav16, Yu17}.  Only one of these planet
candidates, CI Tau b, is around a CTTS \citep[][]{jk16}, offering the
possibility to study a system where a star, its disk, and a massive planet
can interact.  An additional very low mass brown dwarf ($m$sin$i \approx$ 19
$M_{JUP}$) has been discovered around the CTTS AS 205 A \citep[][]{Alm17}. 

\citet{jk16} announced CI Tau b as an $\sim 11~M_{JUP}$ planet in a
possibly eccentric orbit around the CTTS CI Tau.  \citet{biddle18} 
examined the K2 lightcurve of CI Tau and found additional support for the
planetary interpretation of the RV signals observed by \citet{jk16}.
Even more recently, \citet{Cla18} used ALMA to find evidence for 3
additional gas giant planets orbiting CI Tau. 
As an $\sim 2$ Myr old star with fairly mature suite of planets,
this system may have much to reveal about planet formation and migration.

One important parameter of the CI Tau system that has yet to be probed is the 
stellar magnetic field of CI Tau.  Here, we seek to measure the global magnetic
field properties of this star, focusing on the average strength of the
magnetic field at the stellar surface.  Zeeman broadening of K-band
\ion{Ti}{1} lines is an excellent way to measure the magnetic field strength on low mass
young stars \citep[][]{Joh99, Joh07} and has been used
to measure the field strengths of close to 3 dozen young systems to date.
In addition to characterizing the CI Tau magnetic field,
we also compare two methods now used in the literature to measure the
fields of low mass, young stars.  Most studies of the Zeeman
broadening of NIR \ion{Ti}{1} lines follow the analysis described by 
\citet{Joh99} and \citet{Joh07} and use the code SYNTHMAG \citep[][]{Pis99}
to fit a distribution of magnetic field strengths on the stellar
surface.  Recently, \citet{deen13} modified the MOOG \citep{sne73} LTE
atmospheres code to perform radiative transfer in the presence of a 
magnetic field.  \citet{Sok18} used this code to analyze high resolution
NIR spectra of the CTTS TW Hya and measure its mean magnetic field,
finding a value 3.0 kG.  \citet{Yan05} used SYNTHMAG to fit a distribution
of field strengths on the stellar surface finding a mean field strength
of 2.7 kG.  These two studies used spectra taken with different 
instruments at different times, so while it is encouraging that they
find similar mean magnetic fields, the agreement between the two analysis
techniques has not been properly tested.  We seek to perform such a test
in this paper.  The remainder of this paper contains a description of
the observations and data reduction in \S 2, a description of the analysis
in \S3, a discussion of the results in \S 4, and our conclusions are 
presented in \S 5.

\section{Observations and Data Reduction} \label{sec:obs}
\begin{figure}[t!]
\includegraphics*[width=\textwidth,angle=0]{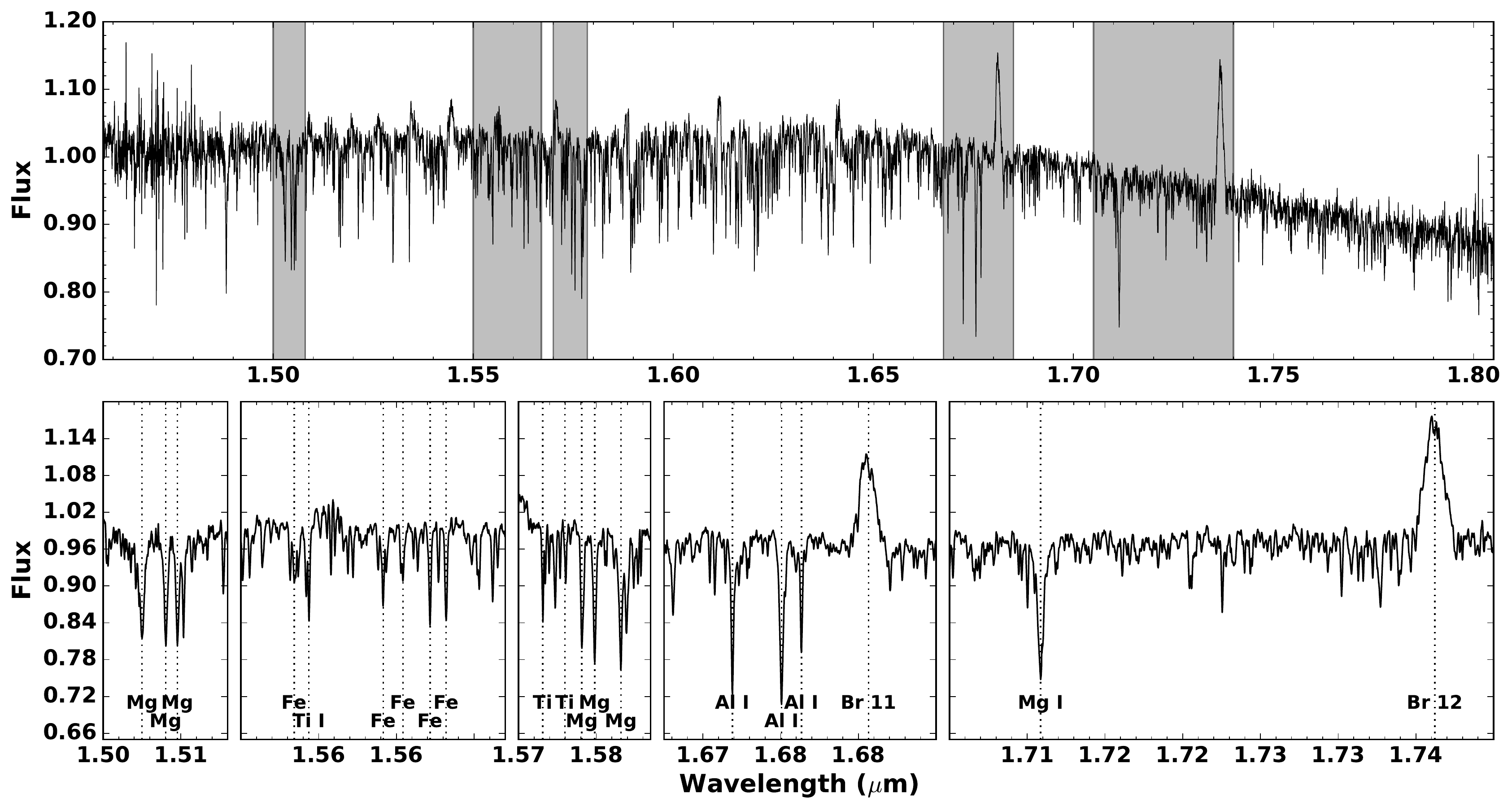}
\caption{\label{fig-spec-H} Views of the final combined IGRINS spectrum of CI Tau. The top panel shows the full H-band spectrum, normalized using the median flux near 1.595$\mu$m in the H-band and 2.192$\mu$m in the K-band. Bottom panels zoom in on regions of interest in the flattened version of the same spectrum.}
\end{figure}

\begin{figure}[h!]
\includegraphics*[width=\textwidth,angle=0]{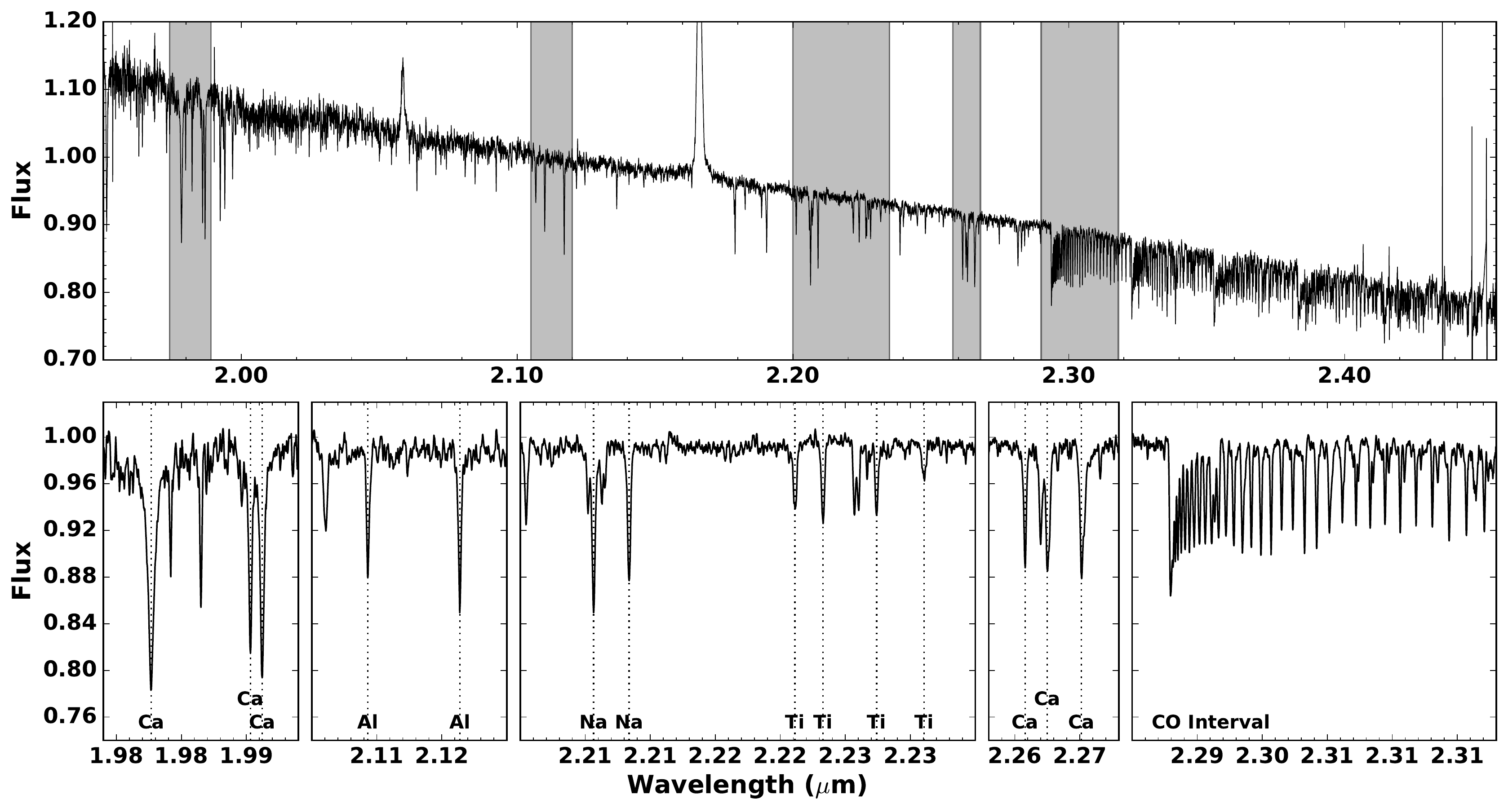}
\caption{\label{fig-spec-K} Same as Figure \ref{fig-spec-H} showing the K-band spectrum of CI Tau. The full K-band spectrum is normalized using the median flux near 2.192$\mu$m.}
\end{figure}

We present a high signal-to-noise IGRINS spectrum of CI Tau, produced by combining IGRINS spectra from 10 separate visits. 
IGRINS (Immersion GRating INfrared Spectrometer) is an extremely powerful instrument that provides a large spectral grasp with high throughput. IGRINS spectra cover the entire H- and K- bands (1.45-2.5 $\mu$m) with a resolving power of R$=\frac{\lambda}{\delta\lambda} =$45,000 \citep{park14,mace16,mace18}. The 10 observations of CI Tau  were obtained with IGRINS on the 4.3m Discovery Channel Telescope of Lowell Observatory over the years of 2016 -- 2017. See Table  \ref{table-obs} for a complete list of the observations. The airmass of CI Tau varied per visit and ranged from 1.02 to 1.275. An A0V telluric standard was observed at a similar airmass either prior or following CI Tau on every night. All targets were nodded along the slit in AB and BA patterns.

To reduce each visits' IGRINS spectroscopic dataset, we use the IGRINS pipeline package \citep[version 2.1 alpha 3; ][]{lee15} to produce a one-dimensional, telluric-corrected spectrum with wavelength solutions derived from OH night sky emission lines at shorter wavelengths and telluric absorption lines  at wavelengths greater than 2.2 $\mu$m. The telluric correction is then performed by dividing the target spectrum by the A0V telluric standard and multiplying by a standard Vega model. The uncertainties of the telluric-corrected spectra are derived by adding the observed uncertainties of the target and standard spectra in quadrature.  In addition to increasing signal, another benefit of combining visits later is that residuals from the telluric correction process are eliminated or greatly reduced, as are most noise elements. Lastly, we correct for the barycenter velocity that corresponds to the Julian time at the middle of each observation.

In preparation for combining the visits, we then  normalize the reduced individual visit spectra and align to a reference frame. We determine velocity shifts between individual visits by cross-correlating each visit with a high signal-to-noise visit as the reference; this reference is marked in Table \ref{table-obs}. 
Because of the potential to artificially broaden the combined spectra, we are very careful and strategic throughout this process.
The cross-correlation is performed across the spectra by 1/4 of an order at a time with 0.1 km/s steps between the reference and the spectrum to be aligned. The observed velocity shift between the two visits then corresponds to the peak of cross-correlation function (for each 1/4 order), which is found by fitting a quadratic to the top 100 points (5 km/s on each side of the peak). The uncertainty of the peak location, i.e. velocity shift,  is estimated from the fitting error found from additionally fitting a Gaussian peak. If the location of the quadratic and Gaussian peaks differ by more than 15 km/s, the solution is thrown out. Therefore, this process also serves as a tool to exclude non-ideal CCF shapes  where the peak value is less reliable. The final velocity shift is found by first taking a 1$\sigma$ cut defined by the standard deviation across all measurements. Then, using the fitting errors as weights, the final velocity shift is computed by a weighted average. Each visit is then shifted to the frame of the reference visit using the measured velocity shift, and interpolated onto the same wavelength solution. Next, the flux of each visit is normalized by dividing by the median flux near 15950$\AA$ in the H-band and 21920$\AA$ in the K-band. 

Finally, the combined spectrum is produced with a weighted average of the 10 aligned visit spectra. The weight for each visit corresponds to the uncertainties of the telluric corrected flux at each pixel. The uncertainties of the combined spectrum are given by the standard deviation of the mean. The final combined spectrum has an average signal-to-noise ratio of $\sim 640$ in the H-band and the K-band.  The full spectra are shown in Figures \ref{fig-spec-H}--\ref{fig-spec-K}, as well as a close look of the highly resolved detail shown by zooming in on some interesting spectral features.

\begin{deluxetable*}{lllll}
\tabletypesize{\footnotesize}
\tablecolumns{8} 
\tablewidth{0pt}
 \tablecaption{IGRINS Observations of CI Tau \label{table-obs}}
 \tablehead{
\colhead{UT Date} &  \colhead{Integration Time} &  \colhead{Observing Sequence} & \colhead{Airmass} & \colhead{A0V Telluric Standard} } 

 \startdata 
20161016	& 300	&	ABBA	&	1.07	&	HIP 21823	\\
20161111\tablenotemark{a}	& 300	&	ABBA	&	1.035	&	HIP 21823	\\
20161112	& 300	&	ABBA x 2	&	1.03	&	HIP 18769	\\
20161115	& 300	&	ABBA	&	1.195	&	HIP 23088	\\
20161125	& 600	&	ABBA x 2	&	1.14	&	HIP 23088	\\
20161126	& 300	&	ABBA x 2	&	1.205	&	HIP 23088	\\
20161208	& 500	&	ABBAAB	&	1.02	&	HIP 21823	\\
20170911	& 180	&	ABBA x 3	&	1.275	&	HIP 23088	\\
20170913	& 300	&	ABBA x 2	&	1.24	&	HIP 23088	\\
20170916	& 300	&	ABBA x 2	&	1.145	&	HIP 23088	\\
\enddata
\tablenotetext{a}{Reference visit}
\end{deluxetable*}

\section{Analysis} \label{sec:analysis}

With this work, we characterize the mean surface magnetic field of the famous young star CI Tau using the combined high-quality IGRINS spectrum. We bolster our results by providing blind, independent measurements obtained with the same IGRINS dataset; at the same time, directly comparing two distinct modeling methods to measure the magnetic field strength via the Zeeman broadening in the NIR.

\begin{figure}[t!]
\includegraphics*[width=0.48\textwidth,angle=0]{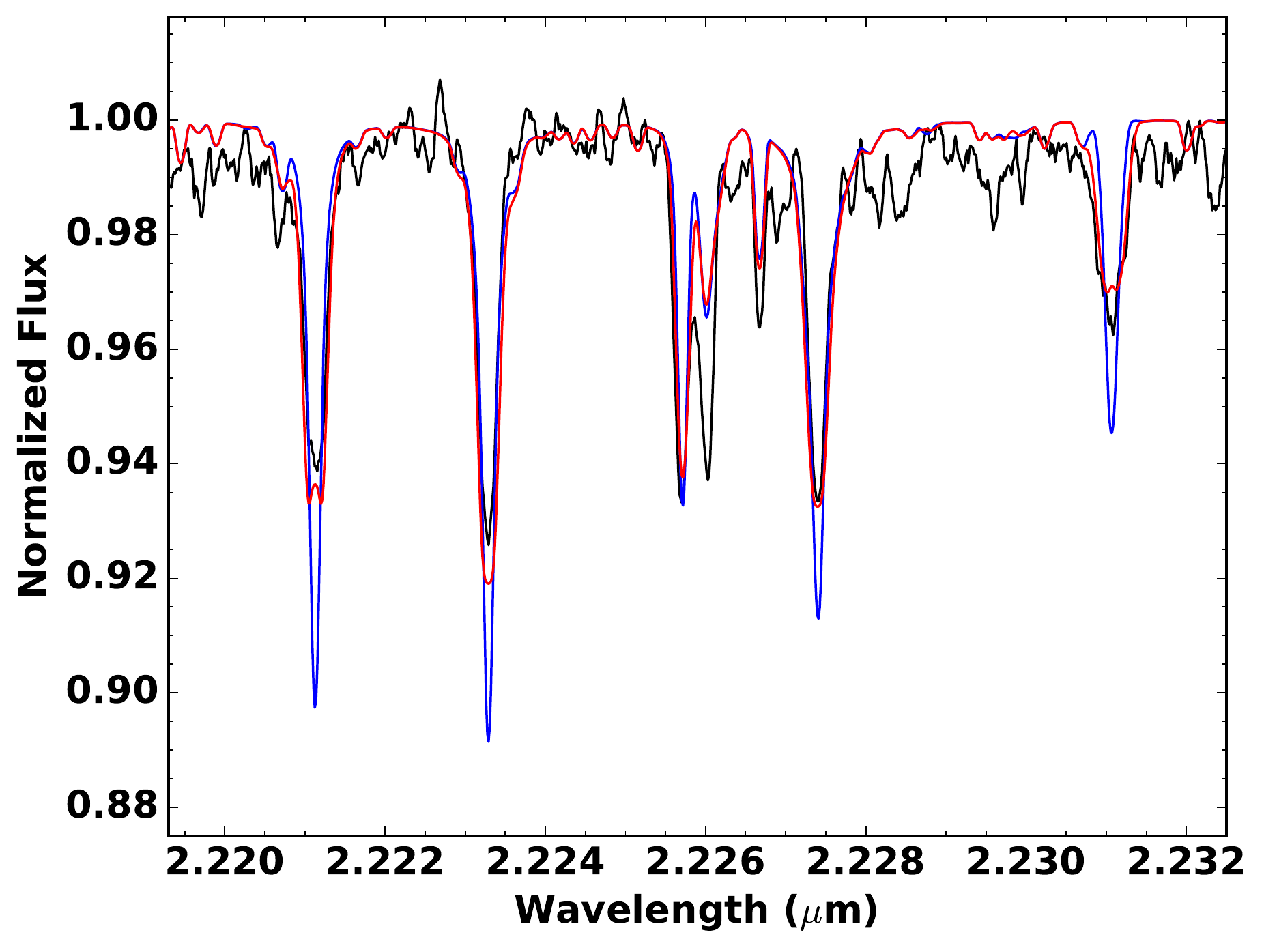}
\includegraphics*[width=0.5\textwidth,angle=0]{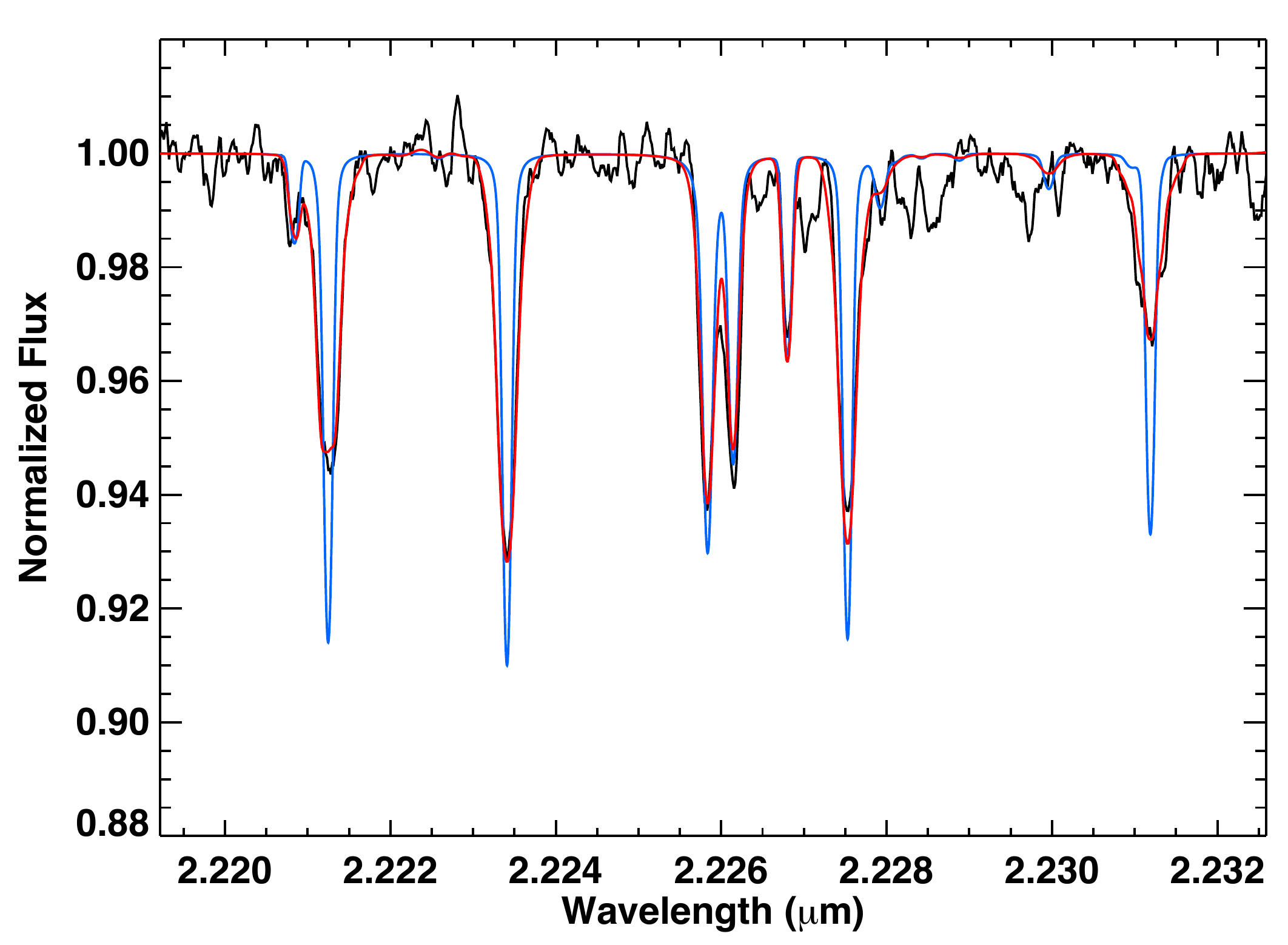}
\caption{\label{fig-fits} A comparison of the results of our two methods to measure the mean magnetic field, illustrated by the best fit models (with and without the magnetic field contribution) in comparison to  the  \ion{Ti}{1} interval of combined IGRINS spectrum of CI Tau. Both methods roughly agree, obtaining a value of $\sim$2.2-2.3 kG, despite different interpretations of the continuum level and method. Left: Method 1/ MoogStokes. Right: Method 2/ SYNTHMAG. Red shows the best fit model corresponding to a mean magnetic field of $B = 2.2-2.3$ kG and the blue line shows the same model, including the same veiling, with the magnetic effects turned off ($B=0.0$ kG).}
\end{figure}

\subsection{Method 1: Magnetic Field Strength  and Stellar Parameterization with MoogStokes} \label{subsec:method_moog}

\subsubsection{Model Grid}

For the first method, we determine the mean magnetic field strength of CI Tau through stellar parameterization with MoogStokes \citep{deen13}. The one-dimensional LTE radiative transfer code Moog \citep{sne73} has been a staple for stellar spectral synthesis since its creation. We optimize on the familiarity and  reliability of this foundation by using a customization of Moog called MoogStokes. MoogStokes synthesizes the emergent spectra of stars including magnetic effects due to a uniform radial magnetic field in the photosphere. MoogStokes calculates the Zeeman splitting of an absorption line by using the spectroscopic terms of the upper and lower state to determine the number, wavelength shift, and polarization of components into which it will split for a given magnetic field strength. Thus, the mean magnetic field strength is one of the fundamental input parameters for MoogStokes.

MoogStokes synthetic spectra are generated using the stellar parameters of effective temperature $T_{\rm eff}$, surface gravity $\log {\rm g} $, and magnetic field strength B. The parameters of effective temperature $T_{\rm eff}$ and surface gravity $\log {\rm g} $ are defined by the model atmospheres, whereas the magnetic field strength can be input as desired. Models are linearly interpolated between grid points as needed. We generate a 3-dimensional grid using solar metallicity \citep[appropriate for YSOs; ][]{padgett96,santos08}  and the MARCS model atmospheres \citep{gus08} resulting in grid that spans $T_{\rm eff}=$ 3000 -- 5000 K, $\log {\rm g}  =$ 3.0 -- 5.0, and $B =$ 0.0 -- 4.0 kG. For each grid model, MoogStokes generates raw emergent spectra synthesized at seven viewing angles; then, it applies the effects of limb darkening and rotational broadening to produce a disk averaged synthetic spectrum \citep{deen13}. We set the rotational broadening to match that of CI Tau, which is  $v \sin i$  $=$ 10.0 km/s \citep{biddle18}. Additionally, we convolve all synthetic spectra with a gaussian kernel to simulate the $R = 45,000$ resolving power of IGRINS.

\subsubsection{Identifying the Best Fit Synthetic Spectrum}

In order to identify the  mean magnetic field strength of CI Tau, we find the best fit MoogStokes synthetic spectrum compared to the combined IGRINS spectrum of CI Tau. We follow a method similar to that of \citet{Sok18}. First, further processing of the combined spectrum is required to compare to the MoogStokes models. We stitch the orders of the observed CI Tau spectrum together and flatten using an interactive python script (based on \url{http://python4esac.github.io/plotting/specnorm.html}). The continuum estimation and flattening process likely contributes one of the greatest sources of uncertainty to the fitting process, and is propagated into estimating the uncertainties (discussed below). 

Throughout the fitting procedure to identify the best fit MoogStokes' synthentic spectral model, we cycle between the stellar parameters (effective temperature, surface gravity, and mean magnetic field) and repeat the process until convergence is reached. For the stellar parameter being investigated, we vary the input value while setting the other two parameters to a constant value. We evaluate the goodness-of-fit across the gridspace \citep[e.g. ][]{Cus08}; then, we adopt this new best value before iterating with the other parameters. The goodness-of-fit is tested over parameter sensitive spectral regions that are similar to \citet{DJ03, Yan05, Sok18}. The effective temperature is evaluated using the Sc and Si lines in the Na interval (2.202-2.212 $\mu$m), the surface gravity with the (2-0) $^{12}$CO interval (2.2925-2.3022 $\mu$m), and the mean magnetic field strength with the \ion{Ti}{1} lines at 2.221$\mu$m, 2.223 $\mu$m, 2.227 $\mu$m, and 2.231 $\mu$m.

To begin the fitting process, we start with an initial model based on the literature: effective temperature of  $T_{\rm eff} =$ 4050 K \citep[log  $T_{\rm eff} =$ 3.6085; ][]{And13,Mcc13}, surface gravity $\log {\rm g} =$ 3.85  extrapolated from CI Tau's stellar mass and age \citep[$M_*=$ 0.8 M$_{\sun}$ at 2 Myr; ][]{Gui14} using the \citet{Bar98} models, and a by-eye estimate for the magnetic field strength of $B = 2.0$ kG. The veiling is measured at 2.2$\mu$m for each synthetic model using a least squares fitting routine to the observed spectrum. The models are artificially veiled using the measured value and a warm dust spectrum corresponding to a $\sim$1500K blackbody.

We find the best fit MoogStokes synthetic spectral model to the combined IGRINS spectrum of CI Tau corresponds to the stellar parameters of $T_{\rm eff} =$ 4025$\pm$25 K, $\log {\rm g} =$ 3.9$\pm$ 0.05, and $B= 2.15 \pm 0.15$ kG with a veiling of r$_{k}=$ 2.3, and plot this comparison in Figure \ref{fig-fits}. Uncertainties on the best fit values of the stellar parameters are estimated by performing a Monte-Carlo simulation, and represent uncertainties in the fitting. We construct simulated observed spectra that are randomly sampled at each pixel from a Gaussian distribution centered on the combined IGRINS flux and with a width based on the uncertainty. We estimate a contribution of an additional 0.5\%  uncertainty to propagate the uncertainties due to the flattening process into the error on our fitting, and add it in quadrature to the existing uncertainties.

\begin{figure}[t!]
\includegraphics*[width=0.5\textwidth,angle=0]{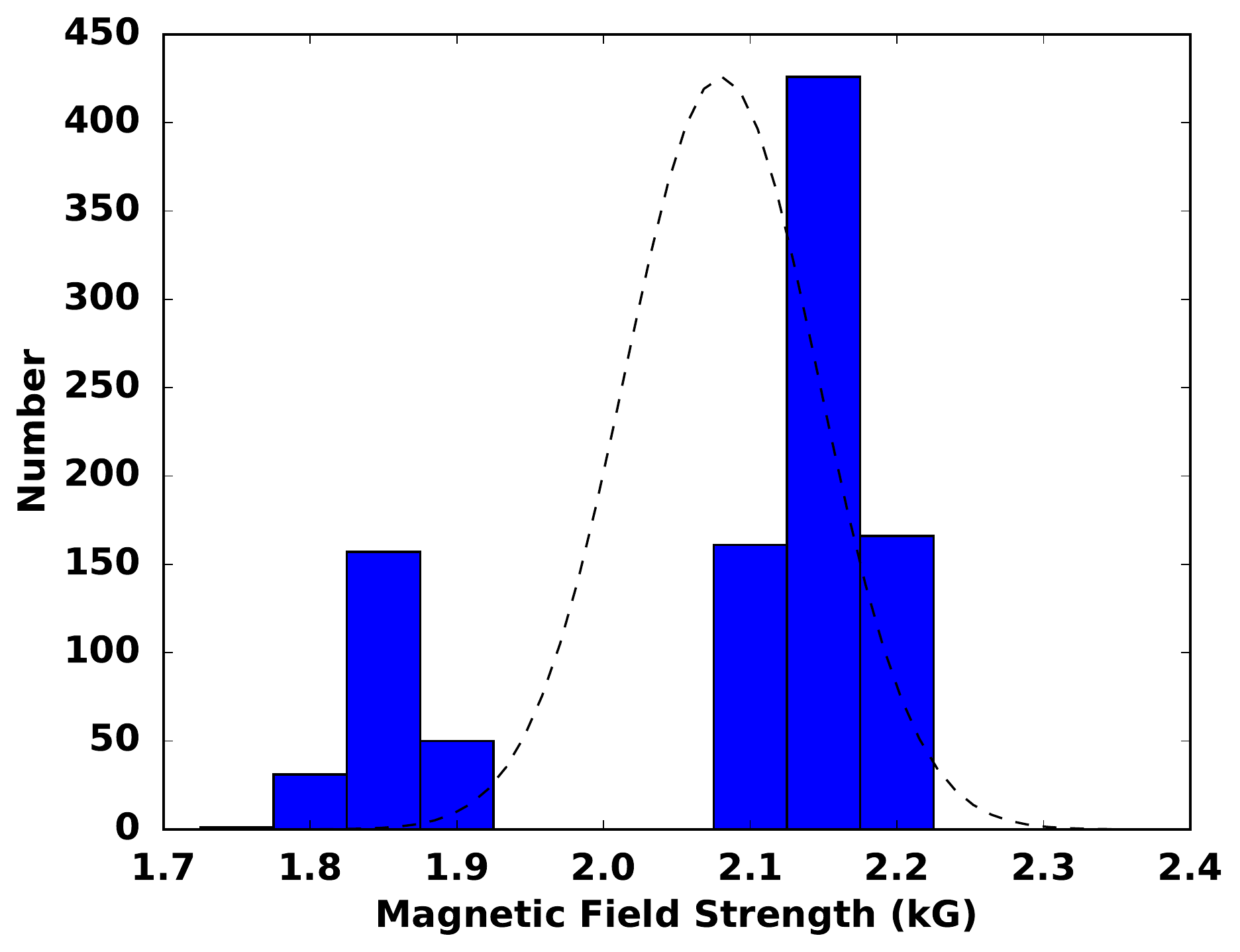}
\caption{\label{fig-bimodal} The distribution of the magnetic field strength values found while running a Monte-Carlo simulation, during which all stellar parameters were allowed to vary, to estimate the uncertainties across the entire parameter space. The final best fit value of the mean magnetic field strength of 2.15 kG corresponds to the peak, mode, and median of this distribution of the best MC fit values  when binned every 0.05 kG as shown.} A Gaussian fit to the distribution is plotted with a dashed line, and the standard deviation gives the adopted uncertainty of $\sim$0.15 kG in the mean field strength measurement. The distribution is clearly bimodal, likely due to different features of the line profiles dominating in the different Monte-Carlo runs, and may suggest that a multi-component magnetic field is more realistic.
\end{figure}

We ran the Monte-Carlo (MC) simulation and fitting routine 1000 times. The best-fitting synthetic spectrum for each randomly-sampled observed spectrum is found by minimizing the goodness-of-fit statistic. We automate the same iterative process as with determining the best fit model, except that the value of the veiling is held constant to the best fit value of r$_{k}$= 2.3.  The standard deviation of the distributions of the best fit MC values for the stellar parameters of effective temperature and surface gravity matched the model grid spacing (25 K and 0.05, respectively). The distribution for the best fit MC values of the magnetic field strength is highly bimodal with the two peaks corresponding to $\sim$1.85 kG and 2.15 kG, as shown in Figure \ref{fig-bimodal}. The mean magnetic field strength of the best fit model corresponds to the peak, mode, and median of this histogram of the best MC fit values. The adopted uncertainty of 0.15 kG is reflective of the standard deviation of the full distribution, which was derived by allowing the stellar parameters to vary in the MC simulation. This bimodality is suggestive of a multi-component magnetic field, which is discussed further in Section \ref{sec-compare}. 

\subsection{Method 2: Magnetic Field Modeling with SYNTHMAG} \label{subsec:method_synthmag}

One of the goals of this study is to perform the magnetic analysis on
CI Tau using two independent methods that have been used for magnetic
analysis of young stars.  The analysis done in this section was performed blindly and thus
independent of knowledge of the results obtained in the last section. 
A number of studies \citep[e.g.][]{Joh04, Yan05, Yan08, Joh09, Yan11} have
analyzed K band spectra of young stars to measure the distribution of magnetic
field strengths on the stellar surface as well as to measure the mean magnetic
field strength using the SYNTHMAG code \citep[][]{Pis99}.  These studies have
generally followed the procedures outlined in \citet{Joh99} and \citet{Joh07}.
We follow these same procedures here.  As such,
the method here and the MoogStokes method outlined in Section \ref{subsec:method_moog} strictly followed their respective methodology; 
we note that  impact of any difference within the processes could be worthy of its own investigation, but is beyond the scope of this project. 
Briefly, these previous studies
utilizing K band spectra with similar spectral resolution to IGRINS have 
found that the best fits to the broadening observed in the \ion{Ti}{1} lines 
result when a distribution of magnetic field strengths is allowed on the
stellar surface.  However, because of the finite spectral resolution and
the intrinsic width of the photospheric line profiles even in the absence of
a magnetic field (e.g. due to thermal, turbulent, and rotational broadening),
only a limited number of different magnetic field components is allowed
when fitting the observed spectra.  It has been found that a 2 kG resolution
in the field results in fairly robust fits - significantly finer resolution
in the allowed magnetic field strengths results in magnetic field
distributions which oscillate fairly substantially due to degeneracies 
associated with trying to constrain field components that very in strength
by a relatively small amount.  Therefore, we fit model spectra to the
observed spectra of CI Tau by allowing field strengths of 0, 2, 4, and 6 kG
and solve for the filling factor of each of these field components.

As noted above, the K band contains several magnetically sensitive \ion{Ti}{1}
lines in addition to a number of relatively magnetically insensitive lines
such as the CO lines of the $v = 2-0$ ro-vibrational transitions near
2.3 $\mu$m.  Historically, for magnetic field analysis we have used the
4 strong \ion{Ti}{1} lines between $2.220 - 2.232$ $\mu$m due to the small 
wavelength grasp of earlier high resolution IR spectrometers such as CSHELL
\citep[][]{Tok90, Gre93} on the NASA IRTF.  IGRINS
contains all 4 of these lines in a single order so they are recorded
simultaneously.  Between the two pairs of \ion{Ti}{1} lines are 4 fairly
strong lines of \ion{Fe}{1}, \ion{Sc}{1}, and \ion{Ca}{1}.  We include these
lines in the analysis here, taking their line data from the VALD atomic
line database \citep[][]{Kup99} and computing their Zeeman splitting patterns
from the transition data contained in the database.  These lines have smaller
Land\'e-$g$ values than the nearby \ion{Ti}{1} lines, but they are non-zero
and are useful for magnetic analysis.

To fit the observed spectrum of CI Tau we compute model spectra with 
SYNTHMAG covering the wavelength range 2.219 - 2.233 $\mu$m and
2.309 - 2.316 $\mu$m (wavelengths given in air).  The first region
contains the magnetically sensitive atomic lines of \ion{Ti}{1} and
the less sensitive \ion{Fe}{1}, \ion{Sc}{1}, and \ion{Ca}{1} lines.
The second region
contains $\sim 10$ CO lines which have very little magnetic sensitivity 
and serve as a check on other line broadening mechanisms such as rotation
and macroturbulence.  In order to perform the spectrum synthesis, basic
atmospheric parameters are required, so we took estimates of these from
the analysis of \citet{Mcc13} who give  $T_{\rm eff} = 4060$ K,
$R_* = 1.41~R_\odot$, and $M_* = 0.80~M_\odot$.  This mass and radius
corresponds to a gravity of $\log {\rm g} = 4.04$.  Since this analysis
was done independent of the MoogStokes analysis above, we selected the
stellar parameters without knowledge of the results of the previous section.
We use the ``next generation"
(NextGen) model atmospheres \citep[][]{All95} to compute the synthetic
spectra.  These model atmospheres are tabulated on a regular grid of 
effective temperature, gravity, and metallicity.  We assume solar 
metallicity for CI Tau and choose the NextGen model from the grid that
most closely matches the stellar parameters from \citet{Mcc13} from above.
Specifically, we take  $T_{\rm eff} = 4000$ K and $\log {\rm g} = 4.0$.  We 
assume a microturbulent broadening of 1 km s$^{-1}$ and a radial-tangential
macroturbulence of 2.0 km s$^{-1}$.  Values for both types of turbulence
are appropriate for a star with CI Tau's parameters \citep[][]{Gra05};
however, our results are quite insensitive to the specific values of
micro- and macroturbulence because other line broadening mechanisms
(rotation and magnetic) dominate.  The last thing needed to compute 
the synthetic spectra is the stellar $v$sin$i$ which we take to be
10.1 km s$^{-1}$ from \citet{biddle18}. We remind the reader that this analysis 
was performed independently from \S \ref{subsec:method_moog}, and therefore 
inputs may vary.

\begin{figure}[t!]
\includegraphics*[width=0.55\textwidth,angle=0]{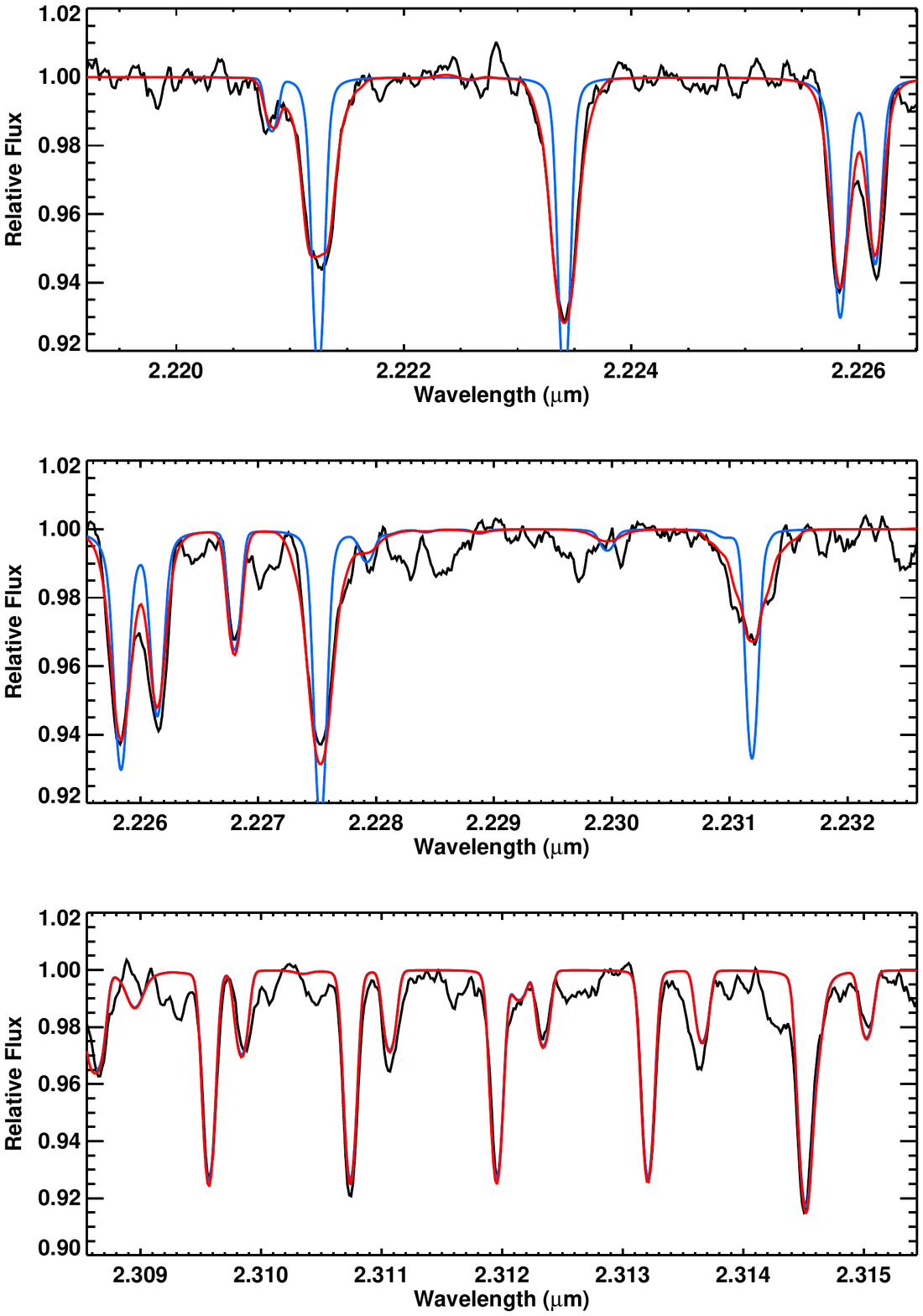}
\includegraphics*[width=0.55\textwidth,angle=0]{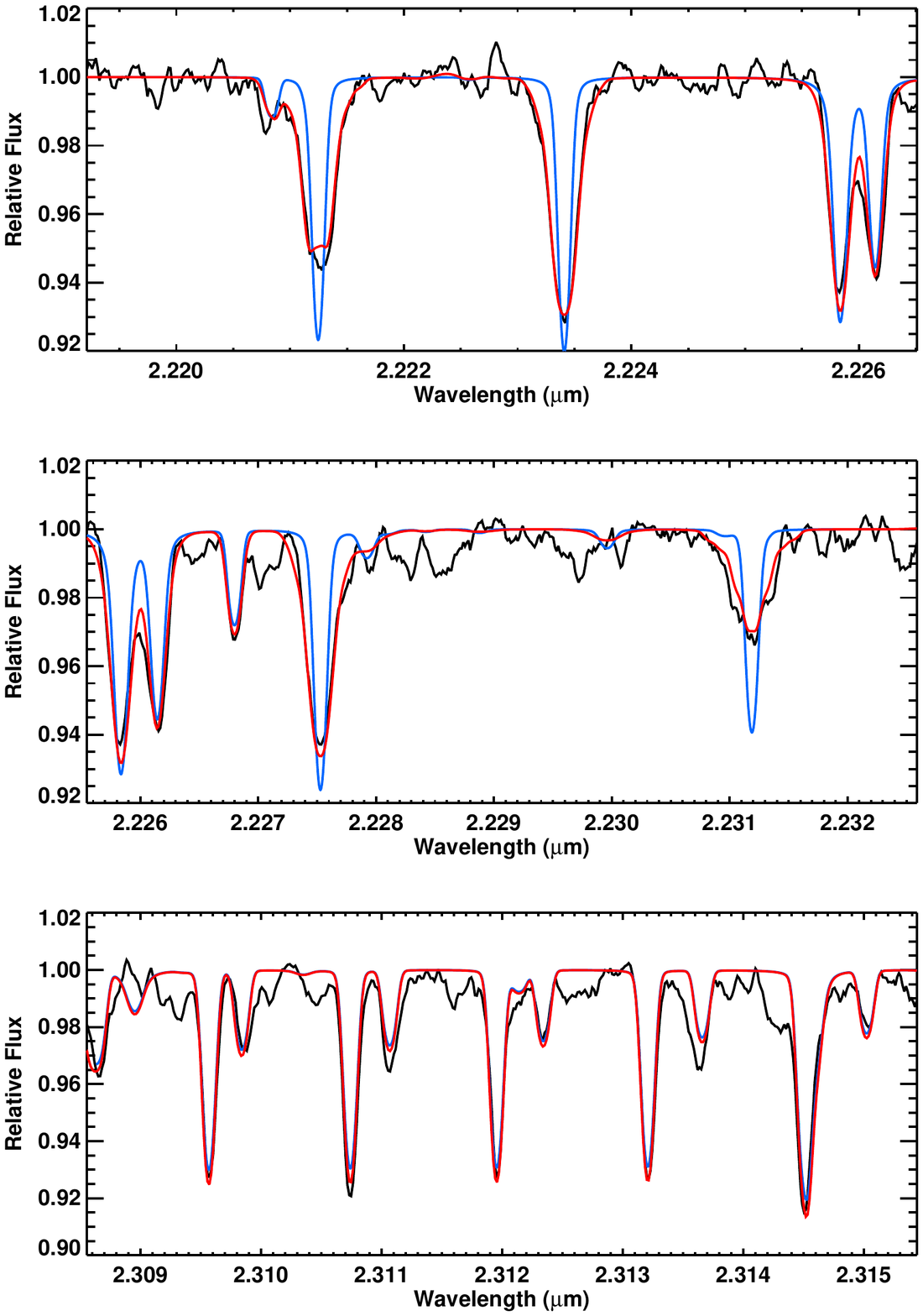}
\caption{\label{fig-fits-synthmag} The results of fitting SYNTHMAG model composed of a distribution of field strengths to the observed CI Tau spectrum.  The IGRINS observations are shown in black in both panels.  In the left set of panels, the $T_{eff} = 4000$
K model with no magnetic field is shown in cyan, and the
model with the magnetic field filling factors for the $T_{eff} = 4000$ K model
from Table 2 is shown in red.  The top two panels show the detail for the
magnetically sensitive \ion{Ti}{1} lines while the bottom panel shows the 
magnetically insensitive CO lines.  Note how narrow the CO lines are and
how broad the \ion{Ti}{1} lines are, indicative of a strong magnetic field.
The right hand panels show the same thing but with the $T_{eff} = 4200$ K
models.}
\end{figure}

As mentioned above, we assume regions on the stellar surface with field
strengths of 0, 2, 4, and 6 kG.  We compute models for each field strength
using SYNTHMAG, assuming the field is oriented radially at the stellar
surface which is generally motivated by solar observations.  In the solar
case, it is also known that the regions of highest photospheric magnetic
field are in dark, cool sunspots.  However, for other stars we do not know
the general relationship between field strength and temperature.  We
therefore assume that each field region has the same temperature (4000 K).
To compute the final profile, we then add the spectra together according to
the assigned filling factor ($f_i$) for each component.  We also add
in veiling from the disk emission in the two regions of spectra we are
fitting.  While these regions are close in wavelength, it is possible the
veiling could be somewhat different between the two, so we allow the 
veiling in each region to be a free parameter.  Finally, we convolve the
resulting profile with a Gaussian to represent the instrumental line
broadening with an assumed spectral resolving power of $R = 45,000$ to match that of IGRINS.  Our
model then has 5 free parameters: filling factors ($f_2$, $f_4$, and $f_6$)
for the 2, 4, and 6 kG field regions ($f_0$ is set by the requirement that
the filling factors sum to 1.0), and the veiling in the two spectral
regions.  It is important to note that the only free parameter in the CO
region is the veiling in this region - all other parameters that affect 
the line strength and width of the CO lines are fixed since they have
negligible magnetic sensitivity.  
We use the the nonlinear least squares 
technique of Marquardt \citep[see][]{Bev92} to fit the observed spectrum by minimizing $\chi^2$ over the regions shown in Figure
\ref{fig-fits-synthmag},
determining the best fit parameters which are given in Table 2. Before comparing the model to the observed spectrum, we normalize the regions of interest in the observed spectrum by dividing 
out a second-order polynomial fit over a small spectral window. Furthermore, this fitting procedure 
is performed on individual orders without requiring any merging.
The
fitted spectral regions and best fit final synthetic spectrum are shown in
left hand panel of Figure \ref{fig-fits-synthmag}. The veiling in the two
regions is found to
be the same within the uncertainties (see next paragraph), so we only report
the mean of the two veilings.  The sum of the field strengths and their 
respective filling factors ($\Sigma Bf$) represents the mean field on the
surface of CI Tau, and we find this value to be $2.26 \pm 0.06$ kG.  We also
plot the best fitting spectrum from this procedure in figure \ref{fig-fits}
for comparison with the results from the MoogStokes analysis.

In order to estimate uncertainties on our fitted field parameters, we again
turn to Monte Carlo simulation.  The data reduction process returns 
uncertainties in the observed spectrum; however, the observed scatter of
the observations around the best fitting model is larger than these estimated
uncertainties.  To provide more realistic values, we compute the standard
deviation (0.0052) of the residuals of the observed spectrum with the fit
subtracted out and add this in quadrature to the uncertainties returned by
the data reduction procedure to get final observational uncertainties.  We then
create a new {\it observed} spectrum by adding Gaussian random noise with an
amplitude given by this revised observational uncertainty to original 
observation, and we analyze this spectrum in the same manner as the original
data.  We repeat this process 1000 times and take the standard deviation of
the resulting values as the uncertainty in the fitted quantities which we
report in Table 2.  For any given fit, the filling factor of the different
field components can trade off each other somewhat, so they have larger
individual uncertainties which are correlated; however, the mean field is
better determined so we only report the uncertainty of this quantity and
the veiling in Table 2.

While the fit in Figure \ref{fig-fits-synthmag} does a good job of fitting the
\ion{Ti}{1} lines, the fit to the \ion{Fe}{1} and \ion{Sc}{1} lines at 2.226
$\mu$m is not quite as good, possibly indicating the atmospheric parameters
(e.g. $T_{eff}$ or log$g$) are not optimally chosen.  Therefore, we repeated
the same analysis but using a NextGen model with $T_{eff} = 4200$ K and
log$g = 4.0$.  This also gives us a chance to see how sensitive our magnetic
results are to an error in the assumed effective temperature.  The results of
this analysis are also reported in Table 2, where it can be seen that we
determine a mean field of $2.30 \pm 0.06$ kG for CI Tau, a value well within
1$\sigma$ of our estimated uncertainty.  This good agreement is likely because
we are detecting actual Zeeman broadening in the \ion{Ti}{1} lines, giving us
a very good handle on the magnetic field properties of this CTTS.  The best
fit model from this fit is shown in the right hand panel of Figure 
\ref{fig-fits-synthmag}.  This model fits the lines at 2.226 $\mu$m somewhat
better, but does not fit the \ion{Ti}{1} lines quite as well, likely
indicating the best temperature for CI Tau is between 4000 and 4200 K, a
result that is not surprising based on previous studies, the potential for an inhomogeneous photosphere, and our analysis
above.

\begin{deluxetable}{ccccccc}
\tabletypesize{\footnotesize}
\tablecolumns{8} 
\tablewidth{0pt}
 \tablecaption{Composite Field Fits \label{table-comp-bfield}}
 \tablehead{
\colhead{Assumed $T_{eff}$} & \colhead{Filling Factor} & \colhead{Filling Factor} & \colhead{Filling Factor} & \colhead{Filling Factor} & \colhead{$\Sigma Bf$} & \colhead{Veiling} \\
\colhead{} &  \colhead{0 kG field} & \colhead{2 kG field} & \colhead{4 kG field} & \colhead{6 kG field} & \colhead{(kG)} & \colhead{}} 
 \startdata 
4000  & 0.21 & 0.54 & 0.17 & 0.08 & $2.26 \pm 0.06$ & $2.01 \pm 0.05$ \\
4200  & 0.19 & 0.56 & 0.16 & 0.09 & $2.30 \pm 0.06$ & $1.83 \pm 0.05$ \\
\enddata
\end{deluxetable}

\section{Discussion}
\subsection{A Comparison of Methods \label{sec-compare}}
A comparison of the results of the blind analysis by our two methods shows good agreement. Both independent methods find the value of the mean magnetic field of CI Tau is $B \sim 2.2$ kG. The resulting best fitting synthetic spectra are shown in Figure \ref{fig-fits}, both with the derived magnetic field and also without any magnetic effects included. Such agreement is not entirely surprising, because the important physics for this result is the same, as with the agreement between different temperature inputs in the SYNTHMAG analysis: the Zeeman broadening of the \ion{Ti}{1} lines caused by a strong magnetic field. The SYNTHMAG method produces a somewhat better fit, which is  not surprising as it is fitting multiple magnetic components. Given that MoogStokes method adopted the goodness-of-fit metric while SYNTHMAG instead adopted $\chi^2$ and a direct comparison of goodness-of-fit between the two methods is complicated by the size of the spectral window used in the respective fits, we can narrow down on the region around the Ti I lines to get an estimate of the difference in the two methods for fitting the magnetic
signatures contained in the data.  Computing $\chi^2$ for the spectral
regions shown in \ref{fig-fits}, we find that the SYNTHMAG fitting results in
a factor of 2.7 improvement in $\chi^2$ relative to MoogStokes.

\begin{deluxetable*}{llllllll}
\tabletypesize{\footnotesize}
\tablewidth{0pt}
 \tablecaption{Compilation of Literature Comparisons of T Tauri Stars \label{table-lit}}
 \tablehead{
\colhead{Source} &  \colhead{T$_{eff}$} &  \colhead{log L/L$\sun$} & \colhead{log L/L$\sun$} & \colhead{Distance} & \colhead{B} & \colhead{Type} & \colhead{References}\\
\colhead{} &  \colhead{[K]} &  \colhead{(literature)} & \colhead{(corrected)} & \colhead{[pc]} & \colhead{[kG]} & \colhead{} & \colhead{(T$_{eff}$,L,B)} } 
\startdata 
AA Tau	&	3792	&	-0.35	&	-0.371	&	136.7	&	2.78	&	cTTS	&	Luh17$+$HH14,HH14,JK07	\\
BP Tau	&	3810	&	-0.38	&	-0.396	&	128.6	&	2.17	&	cTTS	&	Luh17$+$HH14,HH14,JK07	\\
CI Tau	&	4025	&	-0.2	&	-0.095	&	158.0	&	2.2	&	cTTS	&	this work,HH14,this work	\\
CY Tau	&	3515	&	-0.58	&	-0.597	&	128.4	&	1.16	&	cTTS	&	Luh17$+$HH14,HH14,JK07	\\
DE Tau	&	3515	&	-0.28	&	-0.308	&	126.9	&	1.12	&	cTTS	&	Luh17$+$HH14,HH14,JK07	\\
DF Tau	&	3560	&	-0.04	&	-0.142	&	124.5	&	2.9	&	cTTS	&	Luh17$+$HH14,HH14,JK07	\\
DG Tau	&	4020	&	-0.29	&	-0.418	&	120.8	&	2.55	&	cTTS	&	Luh17$+$HH14,HH14,JK07	\\
DH Tau	&	3515	&	-0.66	&	-0.693	&	134.8	&	2.68	&	cTTS	&	Luh17$+$HH14,HH14,JK07	\\
DK Tau	&	3900	&	-0.27	&	-0.347	&	128.1	&	2.64	&	cTTS	&	Luh17$+$HH14,HH14,JK07	\\
DN Tau	&	3846	&	-0.08	&	-0.159	&	127.8	&	0.54	&	cTTS	&	Luh17$+$HH14,HH14,JK07	\\
GG Tau	&	3960	&	0.15	&	0.15	&	140\tablenotemark{b}	&	1.24	&	cTTS	&	Luh17$+$HH14,HH14,JK07	\\
GI Tau	&	3828	&	-0.25	&	-0.314	&	130.0	&	2.73	&	cTTS	&	Luh17$+$HH14,HH14,JK07	\\
GK Tau	&	4068	&	-0.03	&	-0.103	&	128.8	&	2.28	&	cTTS	&	Luh17$+$HH14,HH14,JK07	\\
GM Aur	&	4115	&	-0.31	&	-0.200	&	159.0	&	2.22	&	cTTS	&	Luh17$+$HH14,HH14,JK07	\\
T Tau	&	4870	&	0.85	&	0.831	&	143.7	&	2.37	&	cTTS	&	Luh17$+$HH14,HH14,JK07	\\
Hubble 4	&	3960	&	0.04	&	0.002	&	125.4	&	2.5	&	wTTS	&	Luh17$+$HH14,HH14,JK04	\\
WL 17	&	3400	&	0.255	&	0.203	&	136.5	&	2.9	&	class I	&	D05,D05,JK09	\\
TWA 5A	&	3410	&	-0.61	&	-0.605	&	49.3	&	4.9	&	wTTS	&	Luh17$+$HH14,HH14,Y08	\\
TWA 7	&	3355	&	-0.94	&	-0.940	&	34.0	&	2.3	&	wTTS	&	Luh17$+$HH14,HH14,Y08	\\
TWA 8A	&	3355	&	-0.96	&	-0.897	&	46.2	&	3.3	&	wTTS	&	Luh17$+$HH14,HH14,Y08	\\
TWA 9A	&	4115	&	-0.83	&	-0.410 &	76.2	&	3.5	&	wTTS	&	HH14,HH14,Y08	\\
TWA 9B	&	3322	&	-1.38	&	-1.046	&	76.4	&	3.1	&	wTTS	&	HH14,HH14,Y08	\\
TWA Hya	&	3800	&	-0.72	&	-0.629	&	60.0	&	3	&	cTTS	&	S18,HH14,S18	\\
2M 05353126	&	3669	&	0.184	&	0.184	&	400\tablenotemark{b}	&	2.84	&	cTTS	&	DR16,DR16,Y11	\\
V1227 Ori	&	4200	&	0.086	&	-0.079	&	388.8	&	2.14	&	cTTS	&	Y11,Y11,Y11	\\
2M 05351281	&	3500	&	-0.165	&	-0.296	&	404.4	&	1.7	&	cTTS	&	Y11,Y11,Y11	\\
V1123 Ori	&	3986	&	0.007	&	-0.005	&	394.5	&	2.51	&	wTTS	&	DR16,DR16,Y11	\\
OV Ori	&	4245	&	0.06	&	0.039	&	390.3	&	1.85	&	cTTS	&	DR16,DR16,Y11	\\
V1348 Ori	&	3694	&	-0.101	&	-0.123	&	390.0	&	3.14	&	cTTS	&	DR16,DR16,Y11	\\
LO Ori	&	3600	&	-0.051	&	-0.048	&	401.3	&	3.45	&	cTTS	&	DR16,DR16,Y11	\\
V568 Ori	&	3542	&	-0.061	&	-0.092	&	385.9	&	1.53	&	cTTS	&	DR16,DR16,Y11	\\
LW Ori	&	3961	&	0.136	&	0.142	&	402.8	&	1.3	&	cTTS	&	DR16,DR16,Y11	\\
V1735 Ori	&	4532	&	0.071	&	0.055	&	392.8	&	2.08	&	wTTS	&	DR16,DR16,Y11	\\
V1568 Ori &	3937	&	-0.131	&	-0.131	&	400\tablenotemark{b}	&	1.42	&	wTTS	&	DR16,DR16,Y11	\\
2M 05361049	&	4279	&	-0.02	&	-0.030	&	395.4 &	2.31	&	cTTS	&	DR16,DR16,Y11	\\
2M 05350475	&	3762	&	-0.119	&	-0.132	&	394.3	&	2.79	&	cTTS	&	DR16,DR16,Y11	\\
V1124 Ori	&	3564	&	-0.224	&	-0.326	&	355.7	&	2.09	&	wTTS	&	DR16,DR16,Y11	\\
CHXR 28	&	4060	&	0.08	&	0.283	&	202.1	&	1.5	&	wTTS	&	Lav17,D13,Lav17	\\
YLW 19	&	4590	&	0.12	&	0.199	&	142.3	&	0.8	&	cTTS	&	Lav17,E11,Lav17	\\
KM Ori	&	4730	&	1.0	&	0.954	&	392.5	&	1.9	&	wTTS	&	Lav17,DR12,Lav17	\\
V2062 Oph	&	4730	&	0.3	&	0.164	&	145.3 &	1.8	&	cTTS	&	Lav17,B92,Lav17	\\
\enddata
\tablenotetext{a}{Distances are from Gaia measurements \citep{Bai18}.}
\tablenotetext{b}{Gaia distance not available, and thus the distance used by the luminosity reference is adopted and the luminosity unchanged.}
\tablenotetext{c}{References: B92 $=$ \citet{Bou92}, D05 $=$ \citet{Dop05}, D13 $=$ \citet{Dae13}, DR12 $=$ \citet{Dar12}, DR16 $=$ \citet{Dar16}, E11 $=$ \citet{Eri11}, JK04 $=$ \citet{Joh04}, JK07 $=$ \citet{Joh07}, JK09 $=$ \citet{Joh09}, Lav17 $=$ \citet{Lav17}, Y08 $=$ \citet{Yan08}, Y11 $=$ \citet{Yan11} }
\end{deluxetable*}

Moreover, we find the agreement of our results suggests that measuring the actual Zeeman broadening is quite reliable for estimating the mean magnetic field of the photosphere. Agreement of Zeeman broadening measurements using different models has been found previously as well. However, such agreement was specifically tested here. The list of the differences between the MoogStokes and SYNTHMAG methods is long. Some details that vary are: the assumed continuum in the observed data, the stellar atmosphere used in the code, and even the composition of the magnetic field: a uniform radial field in the MoogStokes method versus a composite field with the SYNTHMAG method.  These fundamental differences in the treatment of the field are similar in nature to the tests performed by \citet{Shu14} who explored two different assumptions regarding the field geometry in their analysis of Zeeman broadening measurements in active M dwarfs, finding very good agreement in the mean field strengths determined under the different assumptions. Thus, it is likely measurements utilizing magnetic broadening are even less sensitive to some of the input parameters as is perhaps expected, at least for strong fields such as in CI Tau. For instance, the continuum value was of great concern in both blind analysis runs -- leading to additional uncertainty being added in the Monte Carlo simulation for propagating this effect into the MoogStokes method results -- and yet the outcome produced very similar measured value regardless of different continuum definitions.

While both methods are successful, relying on essentially the same physics to get at the measurement of the mean magnetic field despite a multitude of differences, the different approaches of the two methods are also their strengths, and why both are interesting to consider and use. MoogStokes, identifies the mean magnetic field strength along with the effective temperature and surface gravity. Thus, this measurement is part of the full picture; the magnetic field strength is not relying on an assumed characterization. Alternatively, the SYNTHMAG analysis employed here fits a multi-component magnetic field that is much more realistic. It identifies not only the mean field strength, but also the filling factors associated with different field components.  The MoogStokes fitting results for CI Tau also may indicate that a single magnetic field strength is not the best description. Particularly, the Monte-Carlo error estimation results in a bimodal distribution for the mean magnetic field strength (Figure \ref{fig-bimodal}). We suspect this is a result of either the broadening in the wings or the more narrow core dominating the fitting process, with the variation between the two resulting from the Monte-Carlo sampling. Thus, the values of the two peaks in this distribution hint that a zero or weak component is an important contribution to the overall field, where this contribution is specifically taken into account in the SYNTHMAG analysis (Table \ref{table-comp-bfield}).

\begin{figure}[t!]
\includegraphics*[width=0.5\textwidth,angle=0]{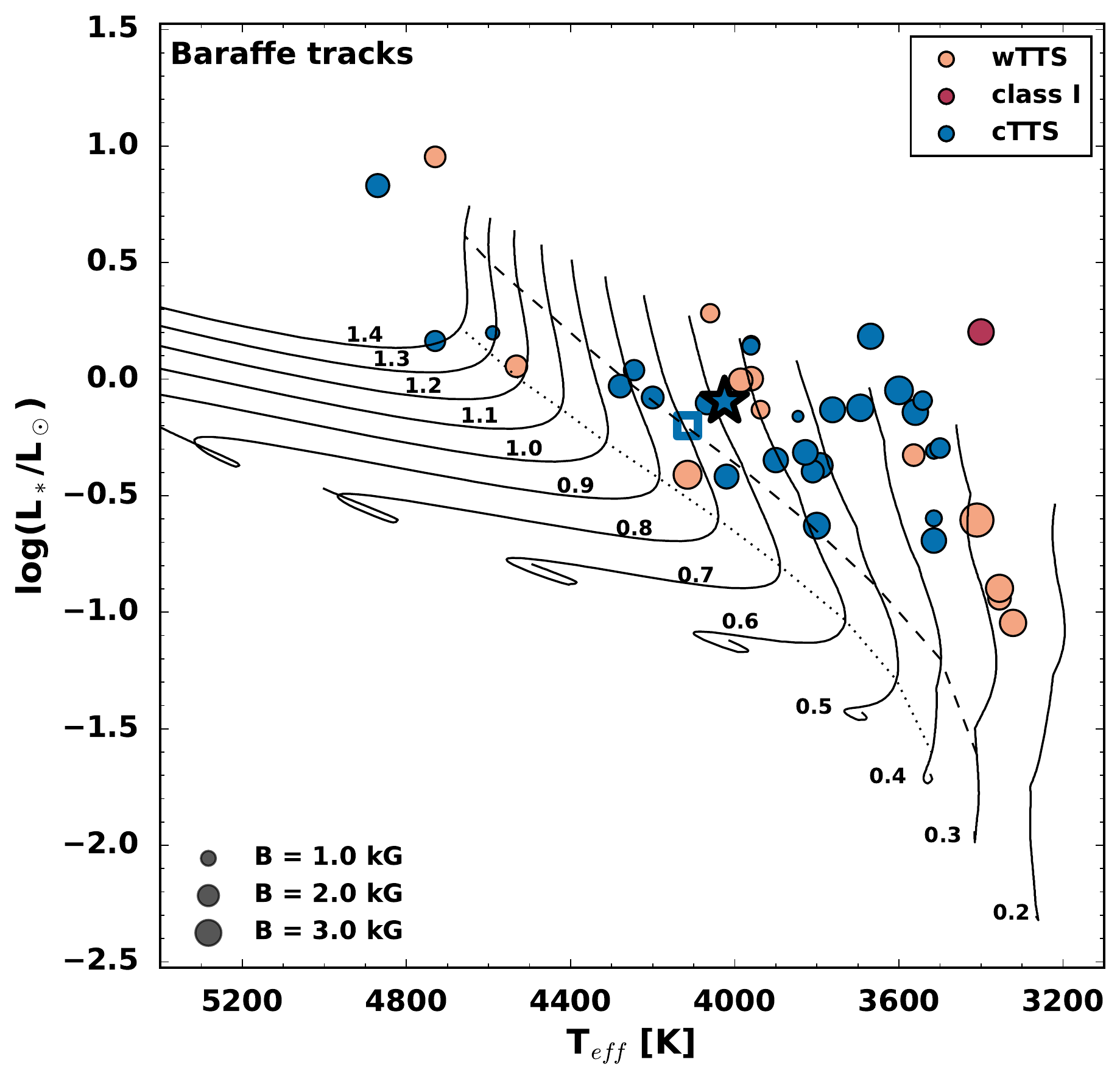}
\includegraphics*[width=0.5\textwidth,angle=0]{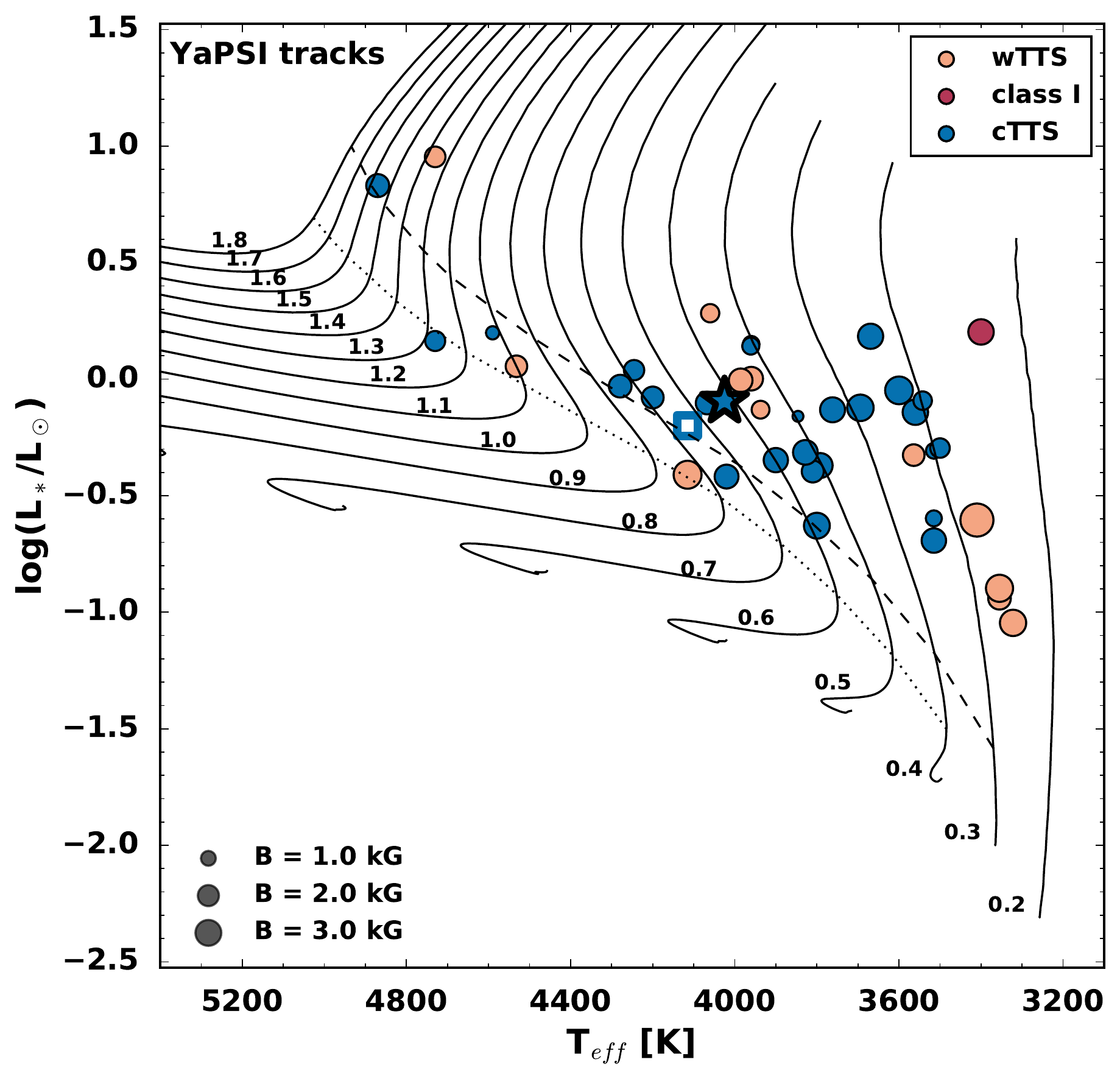}
\caption{\label{fig-hr} CI Tau (as a star symbol) placed in the Hertzsprung-Russell diagram amongst an updated sample of TTSs (circle symbols) for which the surface magnetic field strength is measured from observations of Zeeman broadening. Details and references for the comparison sample are given in Table \ref{table-lit}. Evolutionary tracks of \citet{Bar15,Spa17} are plotted with solid black lines and labeled with the model's stellar mass in Solar masses. The locations on each track at which a radiative core starts forming and the mass of the radiative core is 0.4 the stellar mass are indicated by dashed and dotted lines respectively. The size of the symbols correspond to the strength of the magnetic field. The color of the symbols corresponds to the type of TTS. The square symbol marks GM Aur, which we will use as a reference later.}
\end{figure}

\subsection{CI Tau Amongst Other Magnetic TTSs}

\begin{figure}[t!]
\includegraphics*[width=0.5\textwidth,angle=0]{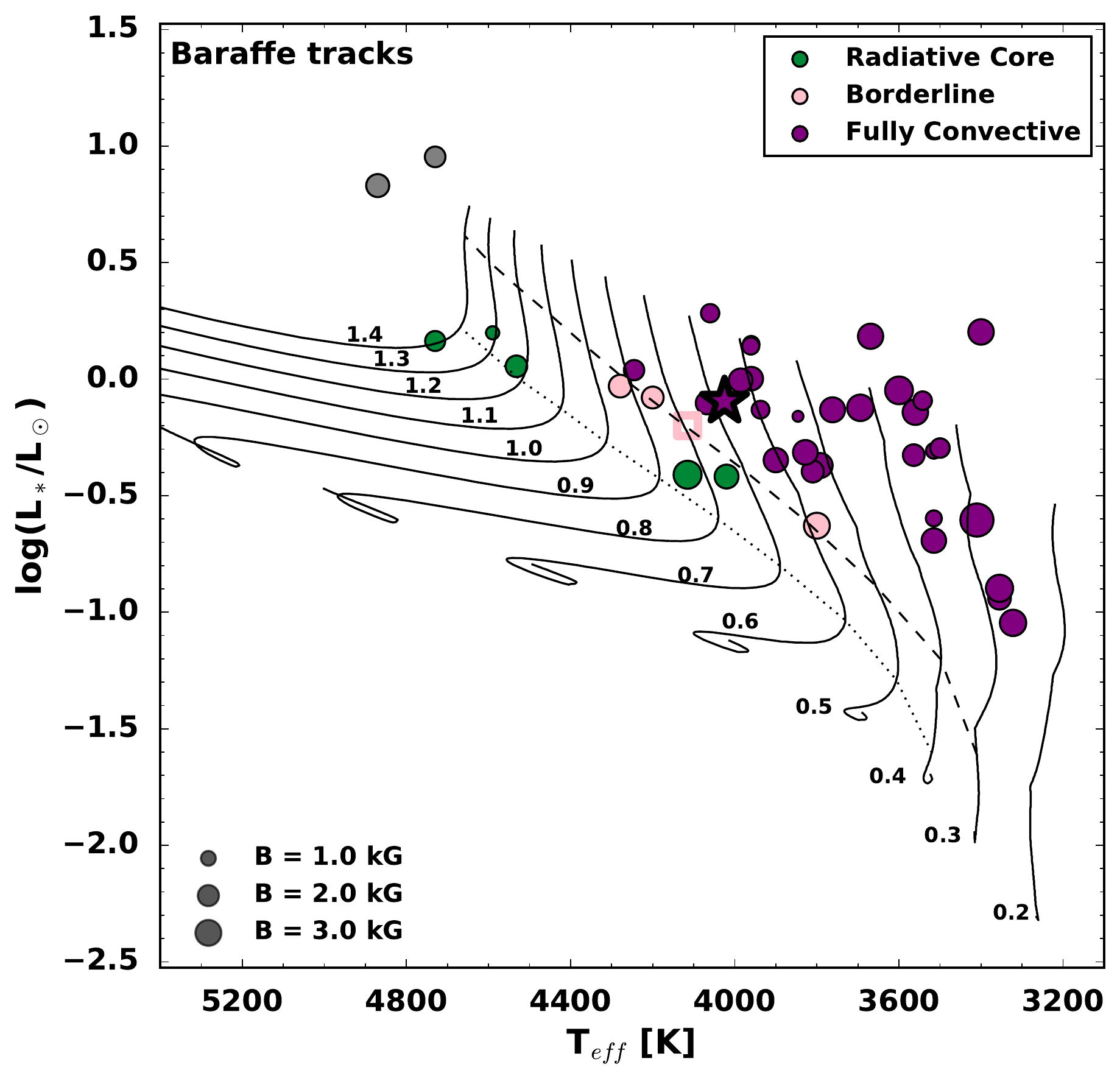}
\includegraphics*[width=0.5\textwidth,angle=0]{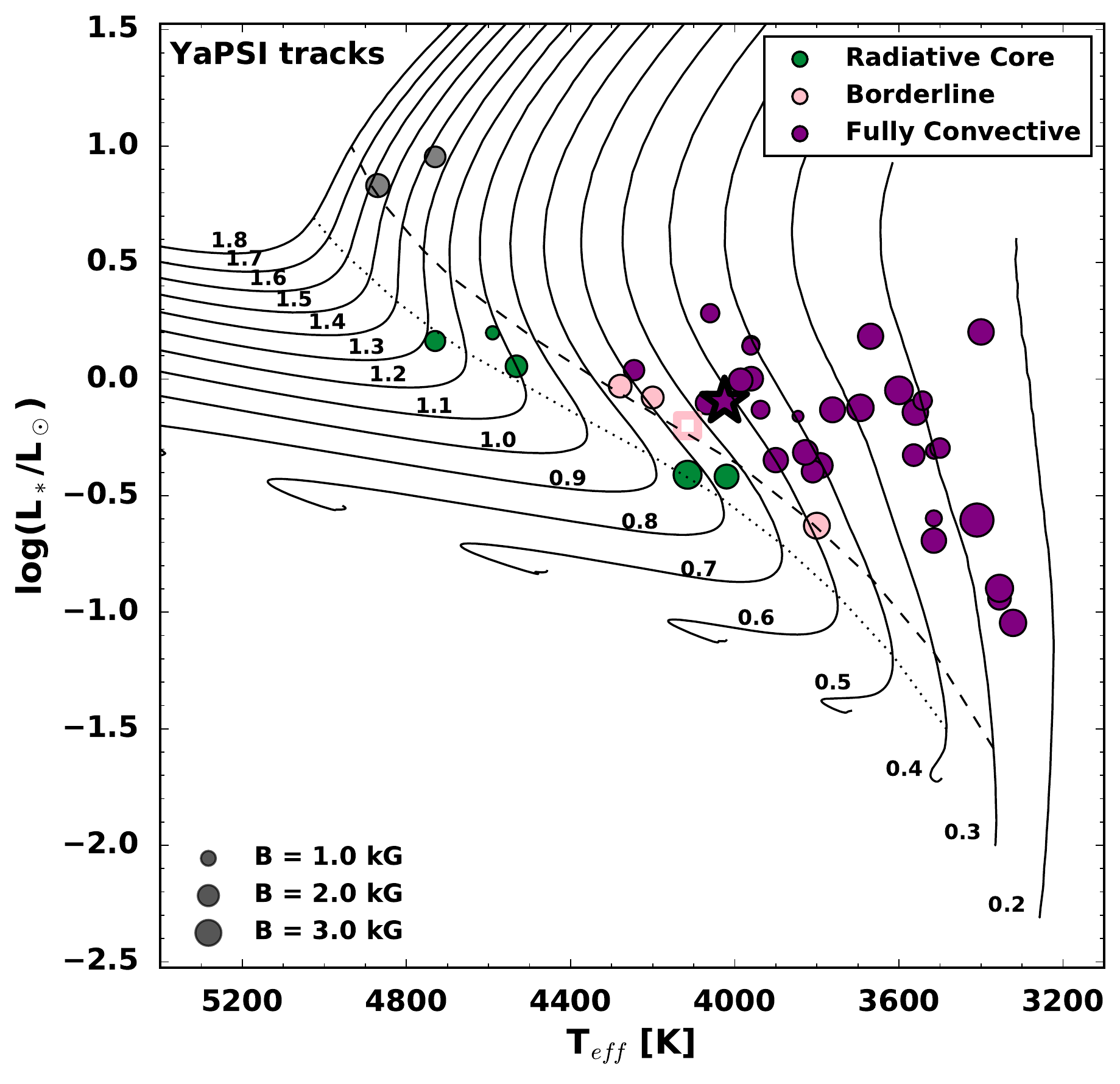}
\caption{\label{fig-hr-evol} An additional versions of the HR diagram shown in Figure \ref{fig-hr}, comparing the evolutionary tracks of \citet{Bar15} and \citet{Spa17} in the left and right panels respectively. Evolutionary tracks are plotted with solid black lines; and the dashed lines indicate the formation of a radiative core and where  the mass of the radiative core is 0.4 the stellar mass. In both panels, the color of the symbols corresponds to the approximate stellar structure of each source. The star symbol marks the location of CI Tau. The square marks GM Aur, which we will use as a reference later in the paper. }
\end{figure}

\begin{figure}[t!]
\includegraphics*[width=0.5\textwidth,angle=0]{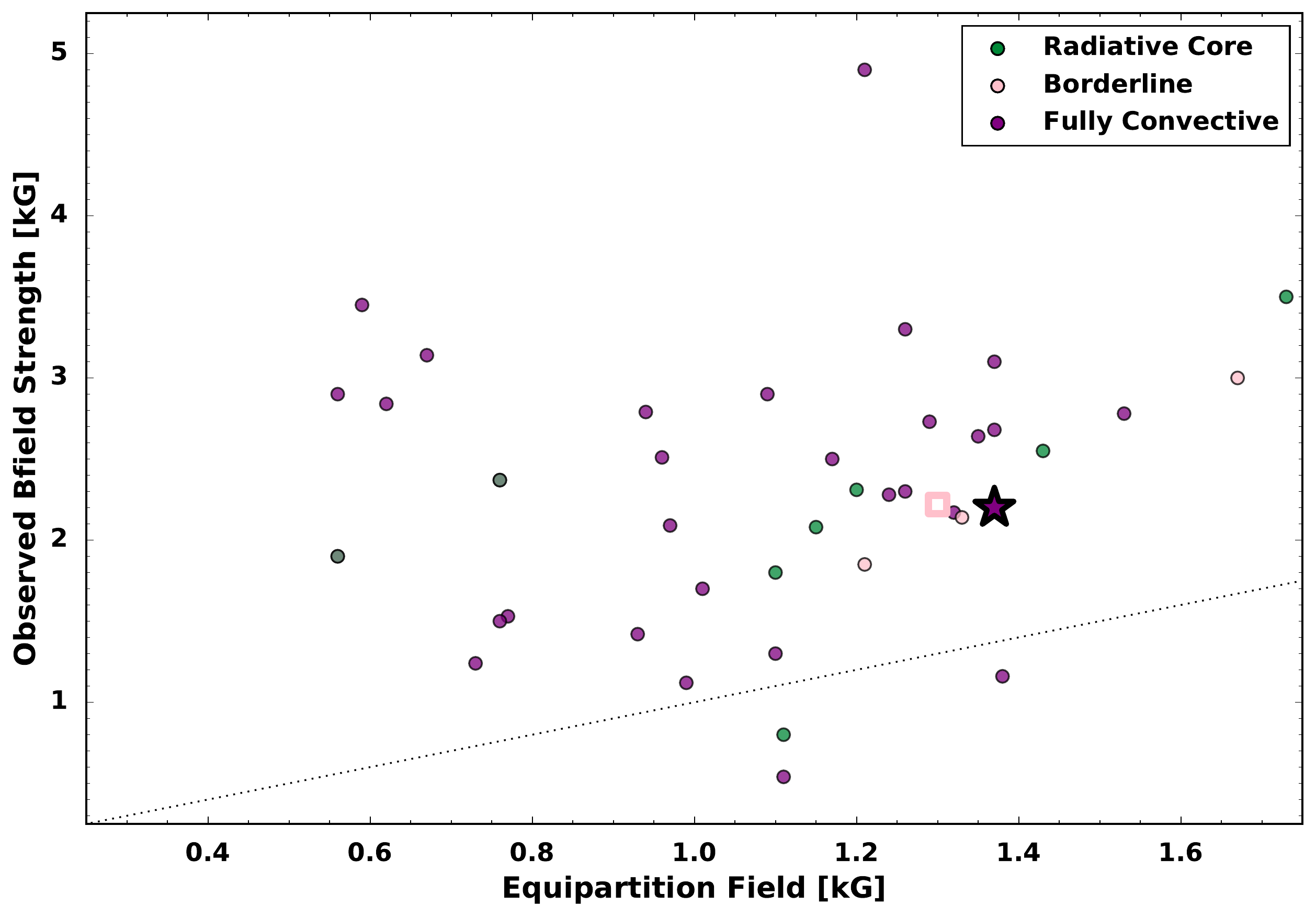}
\caption{\label{fig-equip} The measured mean magnetic field versus the magnetic field predicted by pressure equipartition in the photosphere for the stars in Table \ref{table-lit}.  The dashed line shows the line of equality.  The vast majority of the stars lie above this line indicating their photospheres are magnetically dominated.  In addition, no correlation with the predicted equipartition field exists.}
\end{figure}

\begin{figure}[t!]
\includegraphics*[width=0.5\textwidth,angle=0]{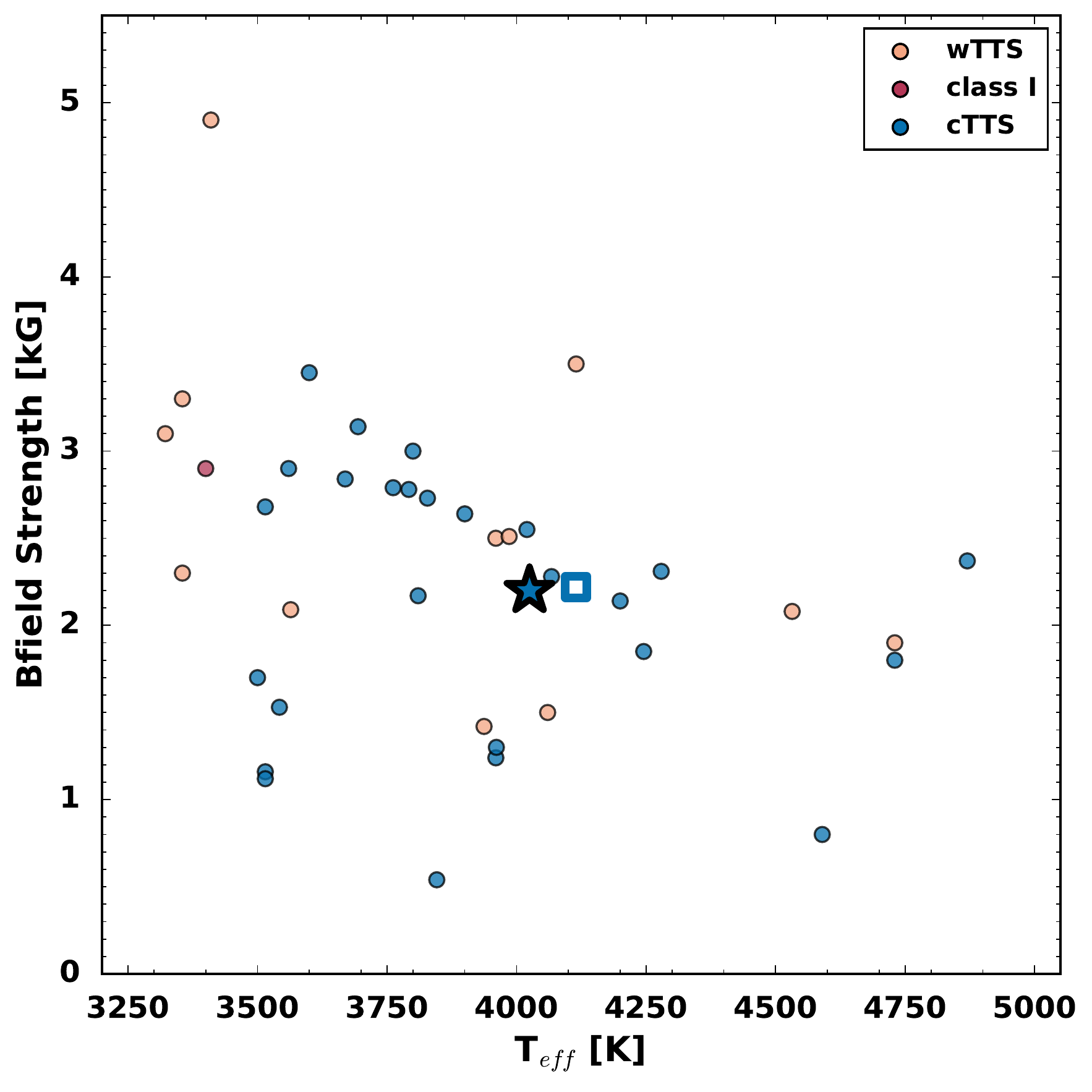}
\includegraphics*[width=0.5\textwidth,angle=0]{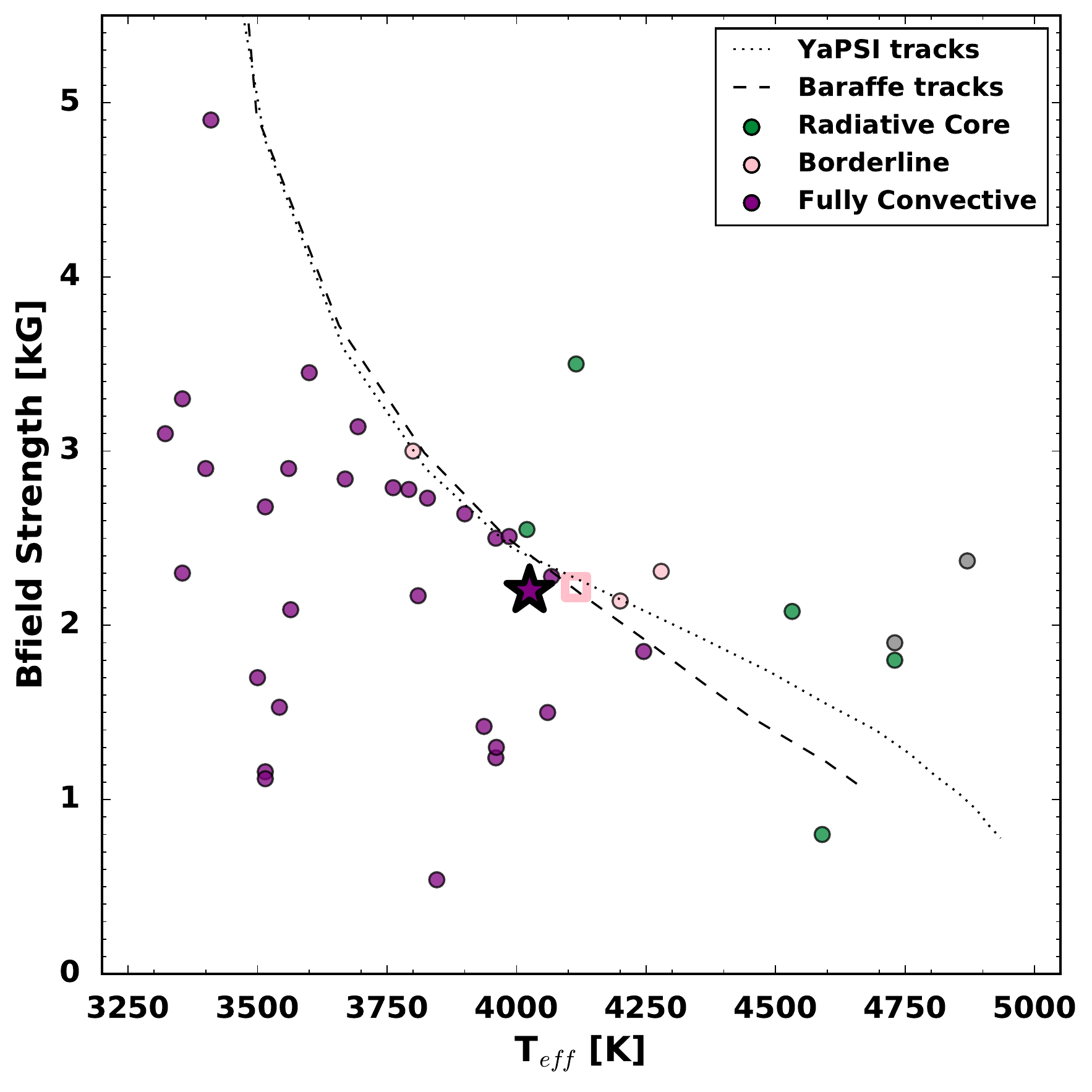}
\caption{\label{fig-b_t} A plot of the mean magnetic field strength versus the effective temperature for CI Tau and a comparative sample of TTSs. The colors are taken from Figure \ref{fig-hr-evol}. A build up of sources is evident near the location of CI Tau (star symbol), roughly seeming following a line with a negative slope. This trend may be indicative of an evolutionary change,} whether it is the conversion to a dynamo from a primordial magnetic field or is explained by invoking the convective boundary. In the case of stellar structure as the origin, we can estimate a scaling relation for the maximum magnetic field strength of an equipartition field for convective instability. Inputing the stellar parameters of the \citet{Bar15} and \citet{Spa17} evolutionary tracks, we plot the maximum magnetic field strength of an equipartition field at the convective boundary (dashed and dotted lines, respectively). This relationship is evaluated at the convective boundary by scaling to the reference TTS GM Aur. Exceptional agreement is shown between the observed data, the inferred stellar structures from the HR diagram, and the predicted boundary shape.
\end{figure}

We put our results into context by placing CI Tau in the Hertzsprung-Russell diagram in Figures \ref{fig-hr}--\ref{fig-hr-evol}. In addition, we directly plot the strength of the magnetic field versus the predictions of the field strength derived from equipartition arguments (Figure \ref{fig-equip}) as well as versus the effective temperature (Figure \ref{fig-b_t}). For comparison, we choose a compilation of TTSs for which the magnetic field strength has been measured through the observed Zeeman broadening; electing for a sample of similar measurements for consistency. This comparison sample includes specific studies of TTSs from Taurus \citep{Joh07}, TW Hydra \citep{Yan08}, and Orion \citep{Yan11} regions, as well as a group of (low to) intermediate TTSs  from \citet{Lav17}. As such, these stars, which are all TTSs, have different ages spanning this young evolutionary phase, and therefore make an excellent test group.  In order to yield a more accurate comparison, we update the stellar parameters of the sample to current literature values whenever possible (see Table \ref{table-lit} for the values and references). Effective temperatures of the TTSs in Taurus and TW Hya are estimated from the spectral typing of \citet{Luh17} using the conversion of \cite{Her14}; effective temperatures of the Orion TTSs are drawn from the IN-SYNC survey using APOGEE \citep{Dar16}. Notably, we correct the stellar luminosities to the current Gaia distance \citep{Bai18}. At the same time, it is important to remember that the magnetic field measurements based on the Zeeman broadening are fairly insensitive to other stellar parameters, such as effective temperature. All values are presented in Table \ref{table-lit}.

It is apparent that CI Tau fits in well with the rest of the observed sample of TTSs  in the HR diagram. CI Tau falls on the vertical Hayashi track, as depicted by the overlaid Baraffe and Yale-Potsdam Stellar Isocrones (YaPSI) evolutionary models \citep[Figure \ref{fig-hr-evol}][]{Bar15,Spa17}, and is in a grouping of other similar stars. As the size of the symbols indicate, this grouping of stars all have similar magnetic field strengths. The HR diagram shown in Figure \ref{fig-hr} is color-coded by TTS type; and the subset of stars near CI Tau include both wTTS and cTTS.  In Figure \ref{fig-hr-evol}, we plot additional HR diagrams to show the evolutionary tracks of \citet{Bar15} and \citet{Spa17} and color code corresponding to the approximate internal structure of the source. In all of the shown HR diagrams, the dashed line indicates the formation of a radiation core, and a dotted line corresponds to where the radiative core contains 40\% of the stellar mass, for the evolutionary tracks being used. CI Tau is clearly above these boundaries and thus likely fully convective. 

Considering the importance of the magnetic field to TTSs and potential planetary systems, it is also worthwhile to examine the observed magnetic field strengths, as there is much to be learned regarding the origin and regulating mechanisms of the field.  An early guiding principle for understanding the magnetic field measurements of cool stars was that the field strength was set by pressure equipartition with the surrounding photosphere.  For active main sequence stars, it was found that the measured fields correlated very well with the equipartition values, and it was also found that the maximum value of the measured field strength equaled the equipartition values to within $\sim 15$\% \citep[]{Saa91,Saa94}.  In Figure \ref{fig-equip} we plot the measured mean field of our TTS sample versus the field predicted by pressure equipartition in the photosphere.  We determine the equipartition field strength by taking the pressure, $P_{eq}$, in the NextGen model atmospheres \citep[]{All95} at the level where the local temperature is equal to the effective temperature for the appropriate effective temperature and gravity of each star.  We then set $P_{eq} = B^2_{eq} / 8\pi$ and solve for the equipartition field, $B_{eq}$.  This represents the maximum field strength that can be confined by the gas pressure in the surrounding non-magnetic atmosphere.  Figure 8 shows no correlation between the measured fields and the predicted fields, with a Spearman rank-order correlation coefficient \citep[]{Zwi00} of 0.20 with an associated false alarm probability of 0.22.  In addition, the vast majority of the measured fields are well above the equipartition values with the median ratio of the observed to equipartition field equal to 1.9.  This suggests that for most TTSs, the entire surface is covered in magnetic field, and therefore, the field strength is not set by pressure equipartition in the visible photosphere.  Similar conclusions were found by \citet{Joh07}.  

 While equipartition does not appear to be operating in this sample, we have found that plotting the  mean magnetic field strength versus the effective temperature produces in an intriguing result. As shown in Figure \ref{fig-b_t}, a trend is apparent in that many of the TTSs appear to lie in a region roughly following a negatively sloped line across the $B$ vs T$_{eff}$ space. This trend is suggestive, and may be indicative of an evolutionary change leading to a pileup in this plot. We find no obvious distinction in the sample according to the type of star (classical or weak-lined TTS). However, color coding the sample by their stellar structure as defined by their respective placement in the HR diagram (Fig. \ref{fig-hr}) may suggest that this trend is related to the evolution and internal structure. We see that the stars that have formed a radiative core lie above and to the right of the  trend defined by the apparent pileup in this plot. 

\begin{figure}[t!]
\includegraphics*[width=0.5\textwidth,angle=0]{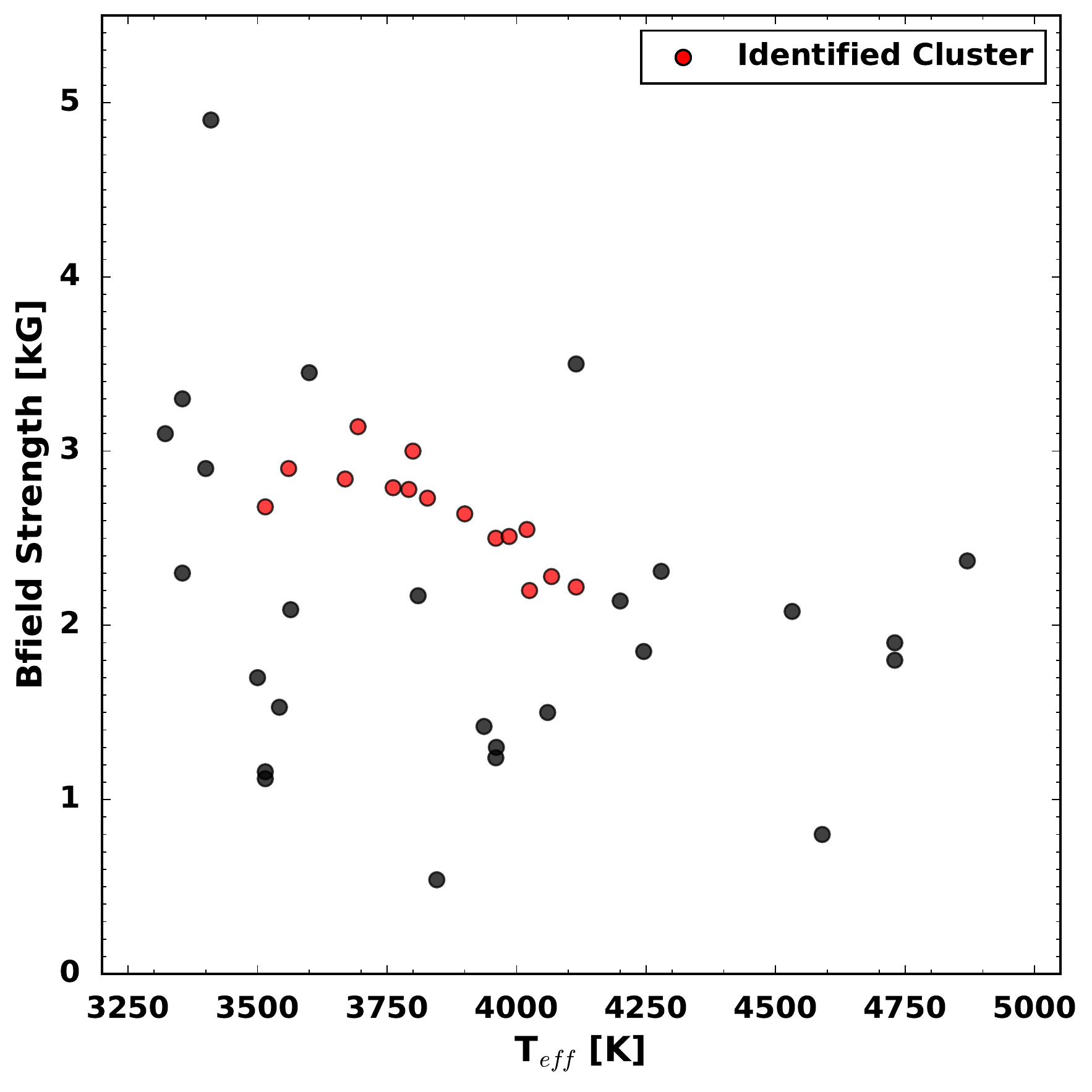}
\caption{\label{fig-trend} A data cluster identified by fitting the DBSCAN clustering algorithm to the literature data, which matches well with the majority of our by-eye identified source pileup. The Spearman correlation coefficient for this pileup group is $\rho = -0.8 $ and is suggestive that the effective temperature and mean magnetic field strength are correlated.}
\end{figure}

Ultimately more data is needed to rigorously test if there is a true pileup observed in this mean magnetic field strength versus effective temperature plot; however, we present an exploratory examination. Statistically, we would not expect a strong correlation coefficient for the entire sample, as only the pileup sources display a by-eye trend. Therefore it is preferable to test the subset of pileup sources. However, identifying this subset can be subjective, especially as not all viewers may even see a pileup.  The presence of a true pileup suggests that some fraction of the points in Figure \ref{fig-b_t} cluster closely in order to define a trend.  Therefore we apply a clustering algorithm to provide an easy and reproducible method of identifying a majority of the pileup group. The effective temperatures and mean magnetic field strength values are first standardized by removing the mean and scaling to unit variance, as is best practice for applying most clustering algorithms. We use DBSCAN \citep[Density-Based Spatial  Clustering of Applications with Noise;][]{Est96} from Python's scikit-learn package \citep{Ped11} as it has the specific advantage of being based on density. DBSCAN can identify any number of arbitrarily sized and shaped data clusters. For the two parameters that define the density, we adopt the default value of 0.5 for the maximum distance to consider two sources to be related, called eps, and set the minimum number of sources to comprise a group to be 20\% of our sample (8 sources).  Reducing the minimum number of sources will result in more, smaller clusters-- and increasing the minimum number of sources $>$8 (and the maximum distance constraint the same) does not lead to any identified clusters in this case . Fitting a DBSCAN model with these inputs results in the identification of one cluster with the rest of the sources labeled as noise. We find that the identified cluster is an excellent match to the trend that is seen by eye and contains the majority of the sources that match our by-eye pileup (see Figure \ref{fig-trend}).

Now that we have identified the potential pileup cluster, we can evaluate the correlation coefficient to better understand the significance of any trend between the effective temperature and mean magnetic field strength. 
We again use the Spearman rank-order correlation coefficient \citep{Zwi00} to test for a relationship between two datasets.
We use a rank-order correlation coefficient since these are designed to test for correlation without assuming any specific underlying functional form for a relationship. As would be expected, the Spearman correlation coefficient $\rho$ is somewhat inconclusive when measured using the entire dataset, with $\rho = -0.33$ with an associated false alarm probability of 0.035. However, the Spearman correlation coefficient for the pileup cluster identified in Figure \ref{fig-trend} is $\rho = -0.82$ with an associated false alarm probability of $1.7 \times 10^{-4}$. A value of $\pm1$ would imply an exact monotonic relationship, and therefor the measured $\rho = -0.82 $ is quite strong, and suggests that the effective temperature and mean magnetic field strength are correlated for the pileup group. 

To try to understand any potential physics involved in the pileup, we can forge a relationship for the magnetic field strength at the fully convective boundary in this parameter space by  invoking the general idea of the equipartition field where the magnetic field pressure is balanced by the thermal gas pressure. Alternatively, \citet{Cha09} discovered that the magnetic field strength of planets and fully convective stars is set by the energy flux. The magnetic pressure is very similar to the magnetic energy density. For this work, we instead investigate a field that represents the maximum field that can be contained by the thermal gas pressure and evaluate it at the convective boundary by scaling to a reference star. We begin with the ideal gas law such that thermal pressure is given by $P_{th}V = N k_B T$ and substitute in for the volume $V = \frac{4}{3} \pi R^3$ and the number of gas molecules $N = \frac{M}{m_p}$. The terms are defined as follows: $P_{th}$ is the thermal pressure, $T$ is the temperature, $k_B$ is the Boltzmann constant, $M$ is the mass of the star, and $m_p$ is the mass of a proton. Then assuming a thermal gas profile such that $P_{th} = \frac{3k_B}{4 \pi m_p} \frac{M}{R^3} T$, we set the thermal pressure equal to the magnetic field pressure of $ P_{mag} = \frac{B^2}{8 \pi}$. While the relations and substitutions made above are fairly simplistic, we use them only to find basic scaling predictions.  This results in the following maximum magnetic field:
\begin{align}
\begin{split}
\frac{6 k_B}{m_p} \times \frac{M} {R^3} \times T= B_{max}^2\\
B_{max} < \left( \frac{\frac{M}{M_*}}{(\frac{R}{R_*})^3} \frac{T}{T_*} \right) ^{1/2}
\end{split}
\end{align}

We can plot this maximum field strength for a fully convective star by inputing the evolutionary track predictions at the convective boundary.  To minimize the impact of assumptions being made and simplify the complex physical relationships,  we can use a reference star and base trends off this reference. The ideal reference star should also be a TTS at the convective boundary. We choose GM Aur from our sample because it is near the fully convective boundary as predicted by both evolutionary models plotted in the HR diagrams (Fig. \ref{fig-hr-evol}). Thus for the reference mass and radius used to approximate the convective boundary in the  $B$ vs T$_{eff}$ plot, we adopt the stellar mass and radius that correspond to the formation of a radiative core in the closest evolutionary track to the position GM Aur in the HR diagrams (model dependent). This choice does impact the vertical placement of this rough boundary, which is most sensitive to the radius. Regardless, we plot the resulting relative maximum magnetic field strength of an equipartition field at the formation of the radiative core in Figure \ref{fig-b_t}. The dashed  and dotted lines are the result of inputting the Baraffe or YaPSI model values at the convective boundary (dashed lines in the HR diagrams) into Equation 1. 
The coherence of the observed data near the plotted boundary is remarkable, especially given the expected uncertainties on the stellar parameters plotted. 
This plot is suggestive that there is likely a physical change in the generation of the magnetic field that leads to the trend see in Figure \ref{fig-b_t} that is related to the stellar structure or age.

The origin of strong magnetic fields for young TTSs is not yet agreed upon. Broadly speaking, the two possible origins for the field are some sort of dynamo action or fossil fields left over from the star formation process. Early inquiries into the equipartition field found that surface convection should be largely suppressed with the large kG magnetic field strengths that are observed, as discussed in \citet{Joh00}, \citet{Joh07}  and above when discussing Figure \ref{fig-equip}. While recent simulations of dynamo action in fully convective stars do not find complete suppression of convection by the field in the simulation zone \citep[e.g.][]{Yad15a, Yad15b}, these simulations are not able to extend all the way to the visible photosphere where the gas pressure decreases dramatically.  Therefore, it is likely the observed large field strengths have a substantial effect on convection in the visible photosphere, even if there is strong convection below the photosphere.  Additionally, there is a lack of clear correlation in TTS magnetic field data with typical dynamo indicators (e.g. rotation period, convective turnover time, Rossby number) that would indicate a dynamo field generation process is active \citep[e.g. ][]{Joh07, Vid14, fol16}.  However, this lack of correlation may be attributed to star-disc interactions or dynamo saturation as with M dwarfs \citep{Rei09}.  Furthermore, while it is generally assumed the magnetic field of M stars are dynamo driven, the strength of the magnetic field in some late M stars is greater than expected from saturation in a standard dynamo \citep[e.g. ][]{Shu17}, similar to what is seen in TTSs.  On the other hand, a recent  non-ideal magnetohydrodynamics simulation shows that a fossil field cannot reproduce the kG magnetic fields that are observed, although resolution may be impacting this result \citep{Wur18}. Further into the TTS evolution, it has been suggested that surface magnetic fields of TTSs could be linked with internal stellar structure. Studies of the magnetic field topology of TTSs, such as the MaPP and MaTYSSE projects \citep[e.g. ][]{Don13,Hil17}, suggest that the magnetic fields become more complex with age and appear to correlate with internal structure. From this view point, \citet{Gre12} proposed that the topology of the magnetic field may even be inferred from the star's location on the HR diagram, assuming a dynamo-generated field. Regardless of the origin or regulating magnetic process, it seems reasonable that there could be an evolutionary pileup of magnetic field measurements if a dynamo is formed or significantly altered at some evolutionary stage, and
that may be what is shown in Figure \ref{fig-b_t}.

Perhaps we are witnessing the conversion from a primordial magnetic field to a dynamo or from one form of dynamo to another (such as a distributed convective dynamo to a solar-like dynamo).  It is quite plausible that the trend seen in Figure \ref{fig-b_t} then represents a maximal efficiency of a convective dynamo; which would explain a pileup at the convective boundary. The most straightforward application goes hand-in-hand with the convective boundary if the trend is caused by a conversion to a solar-like ($\alpha$-$\omega$) dynamo, where the formation of a radiative core could enable the formation of a tachocline. With the development of a core, there is a new source of sheer which will magnify the strength of the magnetic field. Thus a radiative core may boost the magnetic field strength past this rough boundary/trend where the convective limits on the magnetic field no longer hold or are weakened. 


Ultimately, the sample we plot in Figure \ref{fig-b_t} is rather small and therefore might be exaggerating, or even masquerading as,  a trend. A larger sample of sources are needed, particularly at the more evolved stages for lower mass stars, and at less evolved stages for higher mass stars, in order to verify the seemingly clear separation of fully convective versus stars forming a radiative core/ younger versus older stars, as seen with current data. Additionally, as mentioned in the Introduction, the very presence of such strong magnetic fields may alter the estimated effective temperature and placement in the HR diagram when not accounted for.  

\section{Conclusions}

Using IGRINS observations of CI Tau, we present an extremely high signal-to-noise combined spectrum that spans from 1.5 to 2.5 $\mu$m and has a spectral resolving power of $ R = 45,000$. At these NIR wavelengths, the Zeeman effect is enhanced compared to the optical. This broadening is evident in the magnetically sensitive \ion{Ti}{1} lines near 2.2 $\mu$m in the spectrum of CI Tau and is clearly the result of a strong magnetic field present in this young star. We measure the mean surface magnetic field strength of CI Tau to be B$\approx$ 2.25 kG using a blind comparison of two different modeling techniques. 

CI Tau appears to be a perfectly ordinary TTS in the context of this paper. Its mean surface magnetic field strength is similar to other TTSs nearby in the Hertzsprung-Russell diagram. Interestingly, we find that plotting the mean surface magnetic field strength versus the effective temperature for TTSs results in an apparent trend suggestive of some physical change. Whether the observed trend is related to the convective boundary, a switch from primordial to dynamo magnetic fields, coincidence, or something else remains to be determined, and further evidence is needed. Regardless, such findings are promising and the implications for future work is exciting.

\acknowledgments

We thank the anonymous referee for their insightful comments that have improved the paper. KRS would like to thank Dr. Casey Deen for assistance with his MoogStokes code.
This work used the Immersion Grating Infrared Spectrograph (IGRINS) that was developed under a collaboration between the University of Texas at Austin and the Korea Astronomy and Space Science Institute (KASI) with the financial support of the US National Science Foundation under grant AST-1229522 and AST-1702267, of the University of Texas at Austin, and of the Korean GMT Project of KASI.
These results made use of the Discovery Channel Telescope at Lowell Observatory. Lowell is a private, non-profit institution dedicated to astrophysical research and public appreciation of astronomy and operates the DCT in partnership with Boston University, the University of Maryland, the University of Toledo, Northern Arizona University and Yale University. 

This work has made use of data from the European Space Agency (ESA)
mission {\it Gaia} (\url{https://www.cosmos.esa.int/gaia}), processed by
the {\it Gaia} Data Processing and Analysis Consortium (DPAC,
\url{https://www.cosmos.esa.int/web/gaia/dpac/consortium}). Funding
for the DPAC has been provided by national institutions, in particular
the institutions participating in the {\it Gaia} Multilateral Agreement.

%

\vspace{5mm}

\facility{DCT (IGRINS) -- Lowell Observatory's 4.3m Discovery Channel Telescope}


\software{IGRINS pipeline package (version 2.1 alpha 3; Lee \& Gullikson 2016), MoogStokes\citep{deen13}, MARCS \citep{gus08},  SYNTHMAG \citep[][]{Pis99}}





\bibliography{citau}

\begin{thebibliography}{}
\expandafter\ifx\csname natexlab\endcsname\relax\def\natexlab#1{#1}\fi
\providecommand{\url}[1]{\href{#1}{#1}}

\bibitem[{{Aarnio} {et~al.}(2013){Aarnio}, {Matt}, \& {Stassun}}]{Aar13}
{Aarnio}, A.~N., {Matt}, S.~P., \& {Stassun}, K.~G. 2013, Astronomische
  Nachrichten, 334, 77

\bibitem[{{{\'A}d{\'a}mkovics} {et~al.}(2014){{\'A}d{\'a}mkovics}, {Glassgold},
  \& {Najita}}]{Ada14}
{{\'A}d{\'a}mkovics}, M., {Glassgold}, A.~E., \& {Najita}, J.~R. 2014, \apj,
  786, 135

\bibitem[{{Allard} \& {Hauschildt}(1995)}]{All95}
{Allard}, F., \& {Hauschildt}, P.~H. 1995, in The Bottom of the Main Sequence -
  and Beyond, ed. C.~G. {Tinney}, 32

\bibitem[{{Almeida} {et~al.}(2017){Almeida}, {Gameiro}, {Petrov}, {Melo},
  {Santos}, {Figueira}, \& {Alencar}}]{Alm17}
{Almeida}, P.~V., {Gameiro}, J.~F., {Petrov}, P.~P., {et~al.} 2017, \aap, 600,
  A84

\bibitem[{{Andrews} {et~al.}(2013){Andrews}, {Rosenfeld}, {Kraus}, \&
  {Wilner}}]{And13}
{Andrews}, S.~M., {Rosenfeld}, K.~A., {Kraus}, A.~L., \& {Wilner}, D.~J. 2013,
  \apj, 771, 129

\bibitem[{{Bailer-Jones} {et~al.}(2018){Bailer-Jones}, {Rybizki}, {Fouesneau},
  {Mantelet}, \& {Andrae}}]{Bai18}
{Bailer-Jones}, C.~A.~L., {Rybizki}, J., {Fouesneau}, M., {Mantelet}, G., \&
  {Andrae}, R. 2018, \aj, 156, 58

\bibitem[{{Bailey} {et~al.}(2012){Bailey}, {White}, {Blake}, {Charbonneau},
  {Barman}, {Tanner}, \& {Torres}}]{Bai12}
{Bailey}, III, J.~I., {White}, R.~J., {Blake}, C.~H., {et~al.} 2012, \apj, 749,
  16

\bibitem[{{Baraffe} {et~al.}(1998){Baraffe}, {Chabrier}, {Allard}, \&
  {Hauschildt}}]{Bar98}
{Baraffe}, I., {Chabrier}, G., {Allard}, F., \& {Hauschildt}, P.~H. 1998, \aap,
  337, 403

\bibitem[{{Baraffe} {et~al.}(2015){Baraffe}, {Homeier}, {Allard}, \&
  {Chabrier}}]{Bar15}
{Baraffe}, I., {Homeier}, D., {Allard}, F., \& {Chabrier}, G. 2015, \aap, 577,
  A42

\bibitem[{{Barnes} {et~al.}(2001){Barnes}, {Sofia}, \& {Pinsonneault}}]{Bar01}
{Barnes}, S., {Sofia}, S., \& {Pinsonneault}, M. 2001, \apj, 548, 1071

\bibitem[{{Baruteau} {et~al.}(2014){Baruteau}, {Crida}, {Paardekooper},
  {Masset}, {Guilet}, {Bitsch}, {Nelson}, {Kley}, \& {Papaloizou}}]{Bar14}
{Baruteau}, C., {Crida}, A., {Paardekooper}, S.-J., {et~al.} 2014, Protostars
  and Planets VI, 667

\bibitem[{{Bessolaz} {et~al.}(2008){Bessolaz}, {Zanni}, {Ferreira}, {Keppens},
  \& {Bouvier}}]{Bes08}
{Bessolaz}, N., {Zanni}, C., {Ferreira}, J., {Keppens}, R., \& {Bouvier}, J.
  2008, \aap, 478, 155

\bibitem[{{Bevington} \& {Robinson}(1992)}]{Bev92}
{Bevington}, P.~R., \& {Robinson}, D.~K. 1992, {Data reduction and error
  analysis for the physical sciences}

\bibitem[{{Biddle} {et~al.}(2018){Biddle}, {Johns-Krull}, {Llama}, {Prato}, \&
  {Skiff}}]{biddle18}
{Biddle}, L.~I., {Johns-Krull}, C.~M., {Llama}, J., {Prato}, L., \& {Skiff},
  B.~A. 2018, \apjl, 853, L34

\bibitem[{{Bouvier} {et~al.}(2007){Bouvier}, {Alencar}, {Harries},
  {Johns-Krull}, \& {Romanova}}]{Bou07}
{Bouvier}, J., {Alencar}, S.~H.~P., {Harries}, T.~J., {Johns-Krull}, C.~M., \&
  {Romanova}, M.~M. 2007, Protostars and Planets V, 479

\bibitem[{{Bouvier} \& {Appenzeller}(1992)}]{Bou92}
{Bouvier}, J., \& {Appenzeller}, I. 1992, Astronomy and Astrophysics Supplement
  Series, 92, 481

\bibitem[{{Bouvier} {et~al.}(1997){Bouvier}, {Forestini}, \& {Allain}}]{Bou97}
{Bouvier}, J., {Forestini}, M., \& {Allain}, S. 1997, \aap, 326, 1023

\bibitem[{{Calvet} \& {Gullbring}(1998)}]{Cal98}
{Calvet}, N., \& {Gullbring}, E. 1998, \apj, 509, 802

\bibitem[{{Cauley} {et~al.}(2012){Cauley}, {Johns-Krull}, {Hamilton}, \&
  {Lockhart}}]{Cau12}
{Cauley}, P.~W., {Johns-Krull}, C.~M., {Hamilton}, C.~M., \& {Lockhart}, K.
  2012, \apj, 756, 68

\bibitem[{{Chabrier} {et~al.}(2007){Chabrier}, {Gallardo}, \&
  {Baraffe}}]{Cha07}
{Chabrier}, G., {Gallardo}, J., \& {Baraffe}, I. 2007, \aap, 472, L17

\bibitem[{{Chabrier} {et~al.}(2014){Chabrier}, {Johansen}, {Janson}, \&
  {Rafikov}}]{Cha14}
{Chabrier}, G., {Johansen}, A., {Janson}, M., \& {Rafikov}, R. 2014, Protostars
  and Planets VI, 619

\bibitem[{{Chang} {et~al.}(2010){Chang}, {Gu}, \& {Bodenheimer}}]{Cha10}
{Chang}, S.-H., {Gu}, P.-G., \& {Bodenheimer}, P.~H. 2010, \apj, 708, 1692

\bibitem[{{Christensen} {et~al.}(2009){Christensen}, {Holzwarth}, \&
  {Reiners}}]{Cha09}
{Christensen}, U.~R., {Holzwarth}, V., \& {Reiners}, A. 2009, \nat, 457, 167

\bibitem[{{Cieza} \& {Baliber}(2007)}]{Cie07}
{Cieza}, L., \& {Baliber}, N. 2007, \apj, 671, 605

\bibitem[{{Clarke} {et~al.}(2018){Clarke}, {Tazzari}, {Juhasz}, {Rosotti},
  {Booth}, {Facchini}, {Ilee}, {Johns-Krull}, {Kama}, {Meru}, \&
  {Prato}}]{Cla18}
{Clarke}, C.~J., {Tazzari}, M., {Juhasz}, A., {et~al.} 2018, \apjl, 866, L6

\bibitem[{{Crockett} {et~al.}(2012){Crockett}, {Mahmud}, {Prato},
  {Johns-Krull}, {Jaffe}, {Hartigan}, \& {Beichman}}]{Cro12}
{Crockett}, C.~J., {Mahmud}, N.~I., {Prato}, L., {et~al.} 2012, \apj, 761, 164

\bibitem[{{Cushing} {et~al.}(2008){Cushing}, {Marley}, {Saumon}, {Kelly},
  {Vacca}, {Rayner}, {Freedman}, {Lodders}, \& {Roellig}}]{Cus08}
{Cushing}, M.~C., {Marley}, M.~S., {Saumon}, D., {et~al.} 2008, \apj, 678, 1372

\bibitem[{{Da Rio} {et~al.}(2012){Da Rio}, {Robberto}, {Hillenbrand},
  {Henning}, \& {Stassun}}]{Dar12}
{Da Rio}, N., {Robberto}, M., {Hillenbrand}, L.~A., {Henning}, T., \&
  {Stassun}, K.~G. 2012, \apj, 748, 14

\bibitem[{{Da Rio} {et~al.}(2016){Da Rio}, {Tan}, {Covey}, {Cottaar}, {Foster},
  {Cullen}, {Tobin}, {Kim}, {Meyer}, {Nidever}, {Stassun}, {Chojnowski},
  {Flaherty}, {Majewski}, {Skrutskie}, {Zasowski}, \& {Pan}}]{Dar16}
{Da Rio}, N., {Tan}, J.~C., {Covey}, K.~R., {et~al.} 2016, \apj, 818, 59

\bibitem[{{Daemgen} {et~al.}(2013){Daemgen}, {Petr-Gotzens}, {Correia},
  {Teixeira}, {Brandner}, {Kley}, \& {Zinnecker}}]{Dae13}
{Daemgen}, S., {Petr-Gotzens}, M.~G., {Correia}, S., {et~al.} 2013, \aap, 554,
  A43

\bibitem[{{David} {et~al.}(2016){David}, {Hillenbrand}, {Petigura},
  {Carpenter}, {Crossfield}, {Hinkley}, {Ciardi}, {Howard}, {Isaacson}, {Cody},
  {Schlieder}, {Beichman}, \& {Barenfeld}}]{Dav16}
{David}, T.~J., {Hillenbrand}, L.~A., {Petigura}, E.~A., {et~al.} 2016, \nat,
  534, 658

\bibitem[{{Deen}(2013)}]{deen13}
{Deen}, C.~P. 2013, \aj, 146, 51

\bibitem[{{Desort} {et~al.}(2007){Desort}, {Lagrange}, {Galland}, {Udry}, \&
  {Mayor}}]{Des07}
{Desort}, M., {Lagrange}, A.-M., {Galland}, F., {Udry}, S., \& {Mayor}, M.
  2007, \aap, 473, 983

\bibitem[{{Donati} {et~al.}(2013){Donati}, {Gregory}, {Alencar}, {Hussain},
  {Bouvier}, {Jardine}, {M{\'e}nard}, {Dougados}, {Romanova}, \& {MaPP
  Collaboration}}]{Don13}
{Donati}, J.-F., {Gregory}, S.~G., {Alencar}, S.~H.~P., {et~al.} 2013, \mnras,
  436, 881

\bibitem[{{Donati} {et~al.}(2016){Donati}, {Moutou}, {Malo}, {Baruteau}, {Yu},
  {H{\'e}brard}, {Hussain}, {Alencar}, {M{\'e}nard}, {Bouvier}, {Petit},
  {Takami}, {Doyon}, \& {Cameron}}]{Don16}
{Donati}, J.~F., {Moutou}, C., {Malo}, L., {et~al.} 2016, \nat, 534, 662

\bibitem[{{Doppmann} {et~al.}(2005){Doppmann}, {Greene}, {Covey}, \&
  {Lada}}]{Dop05}
{Doppmann}, G.~W., {Greene}, T.~P., {Covey}, K.~R., \& {Lada}, C.~J. 2005, \aj,
  130, 1145

\bibitem[{{Doppmann} \& {Jaffe}(2003)}]{DJ03}
{Doppmann}, G.~W., \& {Jaffe}, D.~T. 2003, \aj, 126, 3030

\bibitem[{{Doppmann} {et~al.}(2003){Doppmann}, {Jaffe}, \& {White}}]{Dop03}
{Doppmann}, G.~W., {Jaffe}, D.~T., \& {White}, R.~J. 2003, \aj, 126, 3043

\bibitem[{{Dullemond} \& {Monnier}(2010)}]{Dul10}
{Dullemond}, C.~P., \& {Monnier}, J.~D. 2010, \araa, 48, 205

\bibitem[{{Edwards} {et~al.}(2006){Edwards}, {Fischer}, {Hillenbrand}, \&
  {Kwan}}]{Edw06}
{Edwards}, S., {Fischer}, W., {Hillenbrand}, L., \& {Kwan}, J. 2006, \apj, 646,
  319

\bibitem[{{Edwards} {et~al.}(1993){Edwards}, {Strom}, {Hartigan}, {Strom},
  {Hillenbrand}, {Herbst}, {Attridge}, {Merrill}, {Probst}, \&
  {Gatley}}]{Edw93}
{Edwards}, S., {Strom}, S.~E., {Hartigan}, P., {et~al.} 1993, \aj, 106, 372

\bibitem[{{Elsner} \& {Lamb}(1977)}]{Els77}
{Elsner}, R.~F., \& {Lamb}, F.~K. 1977, \apj, 215, 897

\bibitem[{{Erickson} {et~al.}(2011){Erickson}, {Wilking}, {Meyer}, {Robinson},
  \& {Stephenson}}]{Eri11}
{Erickson}, K.~L., {Wilking}, B.~A., {Meyer}, M.~R., {Robinson}, J.~G., \&
  {Stephenson}, L.~N. 2011, \aj, 142, 140

\bibitem[{{Ester} {et~al.}(1996){Ester}, {Kriegel}, {Sander}, \& {Xu}}]{Est96}
{Ester}, M., {Kriegel}, H.~P., {Sander}, J., \& {Xu}, X. 1996, in Proceedings
  of the 2nd International Conference on Knowledge Discovery and Data Mining
  (AAAI Press), 226--231

\bibitem[{{Feiden}(2016)}]{Fei16}
{Feiden}, G.~A. 2016, \aap, 593, A99

\bibitem[{{Feiden} \& {Chaboyer}(2013)}]{Fei13}
{Feiden}, G.~A., \& {Chaboyer}, B. 2013, \apj, 779, 183

\bibitem[{{Feiden} \& {Chaboyer}(2014)}]{Fei14}
---. 2014, \apj, 789, 53

\bibitem[{{Folsom} {et~al.}(2016){Folsom}, {Petit}, {Bouvier}, {L{\`e}bre},
  {Amard}, {Palacios}, {Morin}, {Donati}, {Jeffers}, {Marsden}, \&
  {Vidotto}}]{fol16}
{Folsom}, C.~P., {Petit}, P., {Bouvier}, J., {et~al.} 2016, \mnras, 457, 580

\bibitem[{{Gagn{\'e}} {et~al.}(2016){Gagn{\'e}}, {Plavchan}, {Gao},
  {Anglada-Escude}, {Furlan}, {Davison}, {Tanner}, {Henry}, {Riedel},
  {Brinkworth}, {Latham}, {Bottom}, {White}, {Mills}, {Beichman}, {Johnson},
  {Ciardi}, {Wallace}, {Mennesson}, {von Braun}, {Vasisht}, {Prato}, {Kane},
  {Mamajek}, {Walp}, {Crawford}, {Rougeot}, {Geneser}, \&
  {Catanzarite}}]{Gag16}
{Gagn{\'e}}, J., {Plavchan}, P., {Gao}, P., {et~al.} 2016, \apj, 822, 40

\bibitem[{{Gallet} \& {Bouvier}(2013)}]{Gal13}
{Gallet}, F., \& {Bouvier}, J. 2013, \aap, 556, A36

\bibitem[{{Gallet} \& {Bouvier}(2015)}]{Gal15}
---. 2015, \aap, 577, A98

\bibitem[{{Ghosh} \& {Lamb}(1979)}]{Gho79}
{Ghosh}, P., \& {Lamb}, F.~K. 1979, \apj, 232, 259

\bibitem[{{Glassgold} {et~al.}(1997){Glassgold}, {Najita}, \& {Igea}}]{Gla97}
{Glassgold}, A.~E., {Najita}, J., \& {Igea}, J. 1997, \apj, 480, 344

\bibitem[{{Glassgold} {et~al.}(2004){Glassgold}, {Najita}, \& {Igea}}]{Gla04}
---. 2004, \apj, 615, 972

\bibitem[{{Gray}(2005)}]{Gra05}
{Gray}, D.~F. 2005, {The Observation and Analysis of Stellar Photospheres}

\bibitem[{{Greene} {et~al.}(1993){Greene}, {Tokunaga}, {Toomey}, \&
  {Carr}}]{Gre93}
{Greene}, T.~P., {Tokunaga}, A.~T., {Toomey}, D.~W., \& {Carr}, J.~B. 1993, in
  \procspie, Vol. 1946, Infrared Detectors and Instrumentation, ed. A.~M.
  {Fowler}, 313--324

\bibitem[{{Gregory} {et~al.}(2012){Gregory}, {Donati}, {Morin}, {Hussain},
  {Mayne}, {Hillenbrand}, \& {Jardine}}]{Gre12}
{Gregory}, S.~G., {Donati}, J.-F., {Morin}, J., {et~al.} 2012, \apj, 755, 97

\bibitem[{{Guilloteau} {et~al.}(2014){Guilloteau}, {Simon}, {Pi{\'e}tu}, {Di
  Folco}, {Dutrey}, {Prato}, \& {Chapillon}}]{Gui14}
{Guilloteau}, S., {Simon}, M., {Pi{\'e}tu}, V., {et~al.} 2014, \aap, 567, A117

\bibitem[{{Gustafsson} {et~al.}(2008){Gustafsson}, {Edvardsson}, {Eriksson},
  {J{\o}rgensen}, {Nordlund}, \& {Plez}}]{gus08}
{Gustafsson}, B., {Edvardsson}, B., {Eriksson}, K., {et~al.} 2008, \aap, 486,
  951

\bibitem[{{Haisch} {et~al.}(2001){Haisch}, {Lada}, \& {Lada}}]{Hai01}
{Haisch}, Jr., K.~E., {Lada}, E.~A., \& {Lada}, C.~J. 2001, \apjl, 553, L153

\bibitem[{{Hartigan} {et~al.}(1995){Hartigan}, {Edwards}, \&
  {Ghandour}}]{Har95}
{Hartigan}, P., {Edwards}, S., \& {Ghandour}, L. 1995, \apj, 452, 736

\bibitem[{{Hartmann}(1998)}]{Har98}
{Hartmann}, L. 1998, {Accretion Processes in Star Formation}

\bibitem[{{Heller}(2018)}]{Hel18}
{Heller}, R. 2018, ArXiv e-prints, arXiv:1806.06601

\bibitem[{{Herczeg} \& {Hillenbrand}(2014)}]{Her14}
{Herczeg}, G.~J., \& {Hillenbrand}, L.~A. 2014, \apj, 786, 97

\bibitem[{{Hern{\'a}n-Obispo} {et~al.}(2010){Hern{\'a}n-Obispo},
  {G{\'a}lvez-Ortiz}, {Anglada-Escud{\'e}}, {Kane}, {Barnes}, {de Castro}, \&
  {Cornide}}]{Her10}
{Hern{\'a}n-Obispo}, M., {G{\'a}lvez-Ortiz}, M.~C., {Anglada-Escud{\'e}}, G.,
  {et~al.} 2010, \aap, 512, A45

\bibitem[{{Hern{\'a}ndez} {et~al.}(2007){Hern{\'a}ndez}, {Hartmann}, {Megeath},
  {Gutermuth}, {Muzerolle}, {Calvet}, {Vivas}, {Brice{\~n}o}, {Allen},
  {Stauffer}, {Young}, \& {Fazio}}]{Her07}
{Hern{\'a}ndez}, J., {Hartmann}, L., {Megeath}, T., {et~al.} 2007, \apj, 662,
  1067

\bibitem[{{Hill} {et~al.}(2017){Hill}, {Carmona}, {Donati}, {Hussain},
  {Gregory}, {Alencar}, {Bouvier}, \& {Matysse Collaboration}}]{Hil17}
{Hill}, C.~A., {Carmona}, A., {Donati}, J.-F., {et~al.} 2017, \mnras, 472, 1716

\bibitem[{{Hu{\'e}lamo} {et~al.}(2008){Hu{\'e}lamo}, {Figueira}, {Bonfils},
  {Santos}, {Pepe}, {Gillon}, {Azevedo}, {Barman}, {Fern{\'a}ndez}, {di Folco},
  {Guenther}, {Lovis}, {Melo}, {Queloz}, \& {Udry}}]{Hue08}
{Hu{\'e}lamo}, N., {Figueira}, P., {Bonfils}, X., {et~al.} 2008, \aap, 489, L9

\bibitem[{{Ida} \& {Lin}(2010)}]{Ida10}
{Ida}, S., \& {Lin}, D.~N.~C. 2010, \apj, 719, 810

\bibitem[{{Irwin} {et~al.}(2008){Irwin}, {Hodgkin}, {Aigrain}, {Bouvier},
  {Hebb}, {Irwin}, \& {Moraux}}]{Irw08}
{Irwin}, J., {Hodgkin}, S., {Aigrain}, S., {et~al.} 2008, \mnras, 384, 675

\bibitem[{{Johansen} {et~al.}(2014){Johansen}, {Blum}, {Tanaka}, {Ormel},
  {Bizzarro}, \& {Rickman}}]{Joh14}
{Johansen}, A., {Blum}, J., {Tanaka}, H., {et~al.} 2014, Protostars and Planets
  VI, 547

\bibitem[{{Johns-Krull}(2007)}]{Joh07}
{Johns-Krull}, C.~M. 2007, \apj, 664, 975

\bibitem[{{Johns-Krull} \& {Gafford}(2002)}]{Joh02}
{Johns-Krull}, C.~M., \& {Gafford}, A.~D. 2002, \apj, 573, 685

\bibitem[{{Johns-Krull} {et~al.}(2009){Johns-Krull}, {Greene}, {Doppmann}, \&
  {Covey}}]{Joh09}
{Johns-Krull}, C.~M., {Greene}, T.~P., {Doppmann}, G.~W., \& {Covey}, K.~R.
  2009, \apj, 700, 1440

\bibitem[{{Johns-Krull} \& {Valenti}(2000)}]{Joh00}
{Johns-Krull}, C.~M., \& {Valenti}, J.~A. 2000, in Astronomical Society of the
  Pacific Conference Series, Vol. 198, Stellar Clusters and Associations:
  Convection, Rotation, and Dynamos, ed. R.~{Pallavicini}, G.~{Micela}, \&
  S.~{Sciortino}, 371

\bibitem[{{Johns-Krull} {et~al.}(1999){Johns-Krull}, {Valenti}, \&
  {Koresko}}]{Joh99}
{Johns-Krull}, C.~M., {Valenti}, J.~A., \& {Koresko}, C. 1999, \apj, 516, 900

\bibitem[{{Johns-Krull} {et~al.}(2004){Johns-Krull}, {Valenti}, \&
  {Saar}}]{Joh04}
{Johns-Krull}, C.~M., {Valenti}, J.~A., \& {Saar}, S.~H. 2004, \apj, 617, 1204

\bibitem[{{Johns-Krull} {et~al.}(2016){Johns-Krull}, {McLane}, {Prato},
  {Crockett}, {Jaffe}, {Hartigan}, {Beichman}, {Mahmud}, {Chen}, {Skiff},
  {Cauley}, {Jones}, \& {Mace}}]{jk16}
{Johns-Krull}, C.~M., {McLane}, J.~N., {Prato}, L., {et~al.} 2016, \apj, 826,
  206

\bibitem[{{Kaltenegger} {et~al.}(2010){Kaltenegger}, {Eiroa}, {Ribas},
  {Paresce}, {Leitzinger}, {Odert}, {Hanslmeier}, {Fridlund}, {Lammer},
  {Beichman}, {Danchi}, {Henning}, {Herbst}, {L{\'e}ger}, {Liseau}, {Lunine},
  {Penny}, {Quirrenbach}, {R{\"o}ttgering}, {Selsis}, {Schneider}, {Stam},
  {Tinetti}, \& {White}}]{Kal10}
{Kaltenegger}, L., {Eiroa}, C., {Ribas}, I., {et~al.} 2010, Astrobiology, 10,
  103

\bibitem[{{Krishnamurthi} {et~al.}(1997){Krishnamurthi}, {Pinsonneault},
  {Barnes}, \& {Sofia}}]{Kri97}
{Krishnamurthi}, A., {Pinsonneault}, M.~H., {Barnes}, S., \& {Sofia}, S. 1997,
  \apj, 480, 303

\bibitem[{{Kupka} {et~al.}(1999){Kupka}, {Piskunov}, {Ryabchikova}, {Stempels},
  \& {Weiss}}]{Kup99}
{Kupka}, F., {Piskunov}, N., {Ryabchikova}, T.~A., {Stempels}, H.~C., \&
  {Weiss}, W.~W. 1999, \aaps, 138, 119

\bibitem[{{Lagrange} {et~al.}(2013){Lagrange}, {Meunier}, {Chauvin}, {Sterzik},
  {Galland}, {Lo Curto}, {Rameau}, \& {Sosnowska}}]{Lag13}
{Lagrange}, A.-M., {Meunier}, N., {Chauvin}, G., {et~al.} 2013, \aap, 559, A83

\bibitem[{{Lavail} {et~al.}(2017){Lavail}, {Kochukhov}, {Hussain}, {Alecian},
  {Herczeg}, \& {Johns-Krull}}]{Lav17}
{Lavail}, A., {Kochukhov}, O., {Hussain}, G.~A.~J., {et~al.} 2017, \aap, 608,
  A77

\bibitem[{Lee \& Gullikson(2016)}]{lee15}
Lee, J.-J., \& Gullikson, K. 2016, plp: v2.1 alpha 3, , ,
  doi:10.5281/zenodo.56067.
\newblock \url{https://doi.org/10.5281/zenodo.56067}

\bibitem[{{Lin} {et~al.}(1996){Lin}, {Bodenheimer}, \& {Richardson}}]{Lin96}
{Lin}, D.~N.~C., {Bodenheimer}, P., \& {Richardson}, D.~C. 1996, \nat, 380, 606

\bibitem[{{Lin} \& {Papaloizou}(1986)}]{Lin86}
{Lin}, D.~N.~C., \& {Papaloizou}, J. 1986, \apj, 309, 846

\bibitem[{{Luhman} {et~al.}(2017){Luhman}, {Mamajek}, {Shukla}, \&
  {Loutrel}}]{Luh17}
{Luhman}, K.~L., {Mamajek}, E.~E., {Shukla}, S.~J., \& {Loutrel}, N.~P. 2017,
  \aj, 153, 46

\bibitem[{{MacDonald} \& {Mullan}(2009)}]{Mac09}
{MacDonald}, J., \& {Mullan}, D.~J. 2009, \apj, 700, 387

\bibitem[{{Mace} {et~al.}(2016){Mace}, {Kim}, {Jaffe}, {Park}, {Lee}, {Kaplan},
  {Yu}, {Yuk}, {Chun}, {Pak}, {Kim}, {Lee}, {Sneden}, {Afsar}, {Pavel}, {Lee},
  {Oh}, {Jeong}, {Park}, {Kidder}, {Lee}, {Nguyen Le}, {McLane},
  {Gully-Santiago}, {Oh}, {Lee}, {Hwang}, \& {Park}}]{mace16}
{Mace}, G., {Kim}, H., {Jaffe}, D.~T., {et~al.} 2016, in \procspie, Vol. 9908,
  Society of Photo-Optical Instrumentation Engineers (SPIE) Conference Series,
  99080C

\bibitem[{{Mace} {et~al.}(2018){Mace}, {Sokal}, {Lee}, {Oh}, {Park}, {Lee},
  {Good}, {MacQueen}, {Oh}, {Kaplan}, {Kidder}, {Chun}, {Yuk}, {Jeong}, {Pak},
  {Kim}, {Nah}, {Lee}, {Yu}, {Hwang}, {Park}, {Kim}, {Chinn}, {Peck}, {Diaz},
  {Rutten}, {Prato}, {Jacoby}, {Cornelius}, {Hardesty}, {DeGroff}, {Dunham},
  {Levine}, {Nofi}, {Lopez-Valdivia}, {Weinberger}, \& {Jaffe}}]{mace18}
{Mace}, G., {Sokal}, K., {Lee}, J.-J., {et~al.} 2018, in Society of
  Photo-Optical Instrumentation Engineers (SPIE) Conference Series, Vol. 10702,
  Ground-based and Airborne Instrumentation for Astronomy VII, 107020Q

\bibitem[{{Mahmud} {et~al.}(2011){Mahmud}, {Crockett}, {Johns-Krull}, {Prato},
  {Hartigan}, {Jaffe}, \& {Beichman}}]{Mah11}
{Mahmud}, N.~I., {Crockett}, C.~J., {Johns-Krull}, C.~M., {et~al.} 2011, \apj,
  736, 123

\bibitem[{{Matt} \& {Pudritz}(2005)}]{Mat05}
{Matt}, S., \& {Pudritz}, R.~E. 2005, \mnras, 356, 167

\bibitem[{{Matt} {et~al.}(2015){Matt}, {Brun}, {Baraffe}, {Bouvier}, \&
  {Chabrier}}]{Mat15}
{Matt}, S.~P., {Brun}, A.~S., {Baraffe}, I., {Bouvier}, J., \& {Chabrier}, G.
  2015, \apjl, 799, L23

\bibitem[{{Matt} {et~al.}(2010){Matt}, {Pinz{\'o}n}, {de la Reza}, \&
  {Greene}}]{Mat10}
{Matt}, S.~P., {Pinz{\'o}n}, G., {de la Reza}, R., \& {Greene}, T.~P. 2010,
  \apj, 714, 989

\bibitem[{{McClure} {et~al.}(2013){McClure}, {D'Alessio}, {Calvet},
  {Espaillat}, {Hartmann}, {Sargent}, {Watson}, {Ingleby}, \&
  {Hern{\'a}ndez}}]{Mcc13}
{McClure}, M.~K., {D'Alessio}, P., {Calvet}, N., {et~al.} 2013, \apj, 775, 114

\bibitem[{{Mullan} \& {MacDonald}(2001)}]{Mul01}
{Mullan}, D.~J., \& {MacDonald}, J. 2001, \apj, 559, 353

\bibitem[{{Naoz} {et~al.}(2011){Naoz}, {Farr}, {Lithwick}, {Rasio}, \&
  {Teyssandier}}]{Nao11}
{Naoz}, S., {Farr}, W.~M., {Lithwick}, Y., {Rasio}, F.~A., \& {Teyssandier}, J.
  2011, \nat, 473, 187

\bibitem[{{Nguyen} {et~al.}(2012){Nguyen}, {Brandeker}, {van Kerkwijk}, \&
  {Jayawardhana}}]{Ngu12}
{Nguyen}, D.~C., {Brandeker}, A., {van Kerkwijk}, M.~H., \& {Jayawardhana}, R.
  2012, \apj, 745, 119

\bibitem[{{Padgett}(1996)}]{padgett96}
{Padgett}, D.~L. 1996, \apj, 471, 847

\bibitem[{{Papaloizou} {et~al.}(2007){Papaloizou}, {Nelson}, {Kley}, {Masset},
  \& {Artymowicz}}]{Pap07}
{Papaloizou}, J.~C.~B., {Nelson}, R.~P., {Kley}, W., {Masset}, F.~S., \&
  {Artymowicz}, P. 2007, Protostars and Planets V, 655

\bibitem[{{Park} {et~al.}(2014){Park}, {Jaffe}, {Yuk}, {Chun}, {Pak}, {Kim},
  {Pavel}, {Lee}, {Oh}, {Jeong}, {Sim}, {Lee}, {Nguyen Le}, {Strubhar},
  {Gully-Santiago}, {Oh}, {Cha}, {Moon}, {Park}, {Brooks}, {Ko}, {Han}, {Nah},
  {Hill}, {Lee}, {Barnes}, {Yu}, {Kaplan}, {Mace}, {Kim}, {Lee}, {Hwang}, \&
  {Park}}]{park14}
{Park}, C., {Jaffe}, D.~T., {Yuk}, I.-S., {et~al.} 2014, in Society of
  Photo-Optical Instrumentation Engineers (SPIE) Conference Series, Vol. 9147,
  Society of Photo-Optical Instrumentation Engineers (SPIE) Conference Series,
  1

\bibitem[{{Paulson} \& {Yelda}(2006)}]{Pau06}
{Paulson}, D.~B., \& {Yelda}, S. 2006, \pasp, 118, 706

\bibitem[{{Pecaut} {et~al.}(2012){Pecaut}, {Mamajek}, \& {Bubar}}]{Pec12}
{Pecaut}, M.~J., {Mamajek}, E.~E., \& {Bubar}, E.~J. 2012, \apj, 746, 154

\bibitem[{Pedregosa {et~al.}(2011)Pedregosa, Varoquaux, Gramfort, Michel,
  Thirion, Grisel, Blondel, Prettenhofer, Weiss, Dubourg, Vanderplas, Passos,
  Cournapeau, Brucher, Perrot, \& Duchesnay}]{Ped11}
Pedregosa, F., Varoquaux, G., Gramfort, A., {et~al.} 2011, Journal of Machine
  Learning Research, 12, 2825

\bibitem[{{Piskunov}(1999)}]{Pis99}
{Piskunov}, N. 1999, in Astrophysics and Space Science Library, Vol. 243,
  Polarization, ed. K.~N. {Nagendra} \& J.~O. {Stenflo}, 515--525

\bibitem[{{Plavchan} \& {Bilinski}(2013)}]{Pla13}
{Plavchan}, P., \& {Bilinski}, C. 2013, \apj, 769, 86

\bibitem[{{Prato} {et~al.}(2008){Prato}, {Huerta}, {Johns-Krull}, {Mahmud},
  {Jaffe}, \& {Hartigan}}]{Pra08}
{Prato}, L., {Huerta}, M., {Johns-Krull}, C.~M., {et~al.} 2008, \apjl, 687,
  L103

\bibitem[{{Raymond} {et~al.}(2014){Raymond}, {Kokubo}, {Morbidelli},
  {Morishima}, \& {Walsh}}]{Ray14}
{Raymond}, S.~N., {Kokubo}, E., {Morbidelli}, A., {Morishima}, R., \& {Walsh},
  K.~J. 2014, Protostars and Planets VI, 595

\bibitem[{{Rebull}(2001)}]{Reb01}
{Rebull}, L.~M. 2001, \aj, 121, 1676

\bibitem[{{Rebull} {et~al.}(2006){Rebull}, {Stauffer}, {Megeath}, {Hora}, \&
  {Hartmann}}]{Reb06}
{Rebull}, L.~M., {Stauffer}, J.~R., {Megeath}, S.~T., {Hora}, J.~L., \&
  {Hartmann}, L. 2006, \apj, 646, 297

\bibitem[{{Reiners} {et~al.}(2009){Reiners}, {Basri}, \& {Browning}}]{Rei09}
{Reiners}, A., {Basri}, G., \& {Browning}, M. 2009, \apj, 692, 538

\bibitem[{{Romanova} {et~al.}(2009){Romanova}, {Ustyugova}, {Koldoba}, \&
  {Lovelace}}]{Rom09}
{Romanova}, M.~M., {Ustyugova}, G.~V., {Koldoba}, A.~V., \& {Lovelace},
  R.~V.~E. 2009, \mnras, 399, 1802

\bibitem[{{Saar}(1991)}]{Saa91}
{Saar}, S. 1991, in Lecture Notes in Physics, Berlin Springer Verlag, Vol. 380,
  IAU Colloq. 130: The Sun and Cool Stars. Activity, Magnetism, Dynamos, ed.
  I.~{Tuominen}, D.~{Moss}, \& G.~{R{\"u}diger}, 389--400

\bibitem[{{Saar}(1994)}]{Saa94}
{Saar}, S.~H. 1994, in IAU Symposium, Vol. 154, Infrared Solar Physics, ed.
  D.~M. {Rabin}, J.~T. {Jefferies}, \& C.~{Lindsey}, 437

\bibitem[{{Saar} \& {Linsky}(1985)}]{Saa85}
{Saar}, S.~H., \& {Linsky}, J.~L. 1985, \apjl, 299, L47

\bibitem[{{Santos} {et~al.}(2008){Santos}, {Melo}, {James}, {Gameiro},
  {Bouvier}, \& {Gomes}}]{santos08}
{Santos}, N.~C., {Melo}, C., {James}, D.~J., {et~al.} 2008, \aap, 480, 889

\bibitem[{{Segura} {et~al.}(2010){Segura}, {Walkowicz}, {Meadows}, {Kasting},
  \& {Hawley}}]{Seg10}
{Segura}, A., {Walkowicz}, L.~M., {Meadows}, V., {Kasting}, J., \& {Hawley}, S.
  2010, Astrobiology, 10, 751

\bibitem[{{Setiawan} {et~al.}(2008){Setiawan}, {Henning}, {Launhardt},
  {M{\"u}ller}, {Weise}, \& {K{\"u}rster}}]{Set08}
{Setiawan}, J., {Henning}, T., {Launhardt}, R., {et~al.} 2008, \nat, 451, 38

\bibitem[{{Setiawan} {et~al.}(2007){Setiawan}, {Weise}, {Henning}, {Launhardt},
  {M{\"u}ller}, \& {Rodmann}}]{Set07}
{Setiawan}, J., {Weise}, P., {Henning}, T., {et~al.} 2007, \apjl, 660, L145

\bibitem[{{Shu} {et~al.}(1994){Shu}, {Najita}, {Ostriker}, {Wilkin}, {Ruden},
  \& {Lizano}}]{Shu94}
{Shu}, F., {Najita}, J., {Ostriker}, E., {et~al.} 1994, \apj, 429, 781

\bibitem[{{Shulyak} {et~al.}(2017){Shulyak}, {Reiners}, {Engeln}, {Malo},
  {Yadav}, {Morin}, \& {Kochukhov}}]{Shu17}
{Shulyak}, D., {Reiners}, A., {Engeln}, A., {et~al.} 2017, Nature Astronomy, 1,
  0184

\bibitem[{{Shulyak} {et~al.}(2014){Shulyak}, {Reiners}, {Seemann}, {Kochukhov},
  \& {Piskunov}}]{Shu14}
{Shulyak}, D., {Reiners}, A., {Seemann}, U., {Kochukhov}, O., \& {Piskunov}, N.
  2014, \aap, 563, A35

\bibitem[{{Skumanich}(1972)}]{Sku72}
{Skumanich}, A. 1972, \apj, 171, 565

\bibitem[{{Sneden}(1973)}]{sne73}
{Sneden}, C.~A. 1973, PhD thesis, THE UNIVERSITY OF TEXAS AT AUSTIN.

\bibitem[{{Sokal} {et~al.}(2018){Sokal}, {Deen}, {Mace}, {Lee}, {Oh}, {Kim},
  {Kidder}, \& {Jaffe}}]{Sok18}
{Sokal}, K.~R., {Deen}, C.~P., {Mace}, G.~N., {et~al.} 2018, \apj, 853, 120

\bibitem[{{Spada} {et~al.}(2017){Spada}, {Demarque}, {Kim}, {Boyajian}, \&
  {Brewer}}]{Spa17}
{Spada}, F., {Demarque}, P., {Kim}, Y.-C., {Boyajian}, T.~S., \& {Brewer},
  J.~M. 2017, \apj, 838, 161

\bibitem[{{Stassun} {et~al.}(2001){Stassun}, {Mathieu}, {Vrba}, {Mazeh}, \&
  {Henden}}]{Sta01}
{Stassun}, K.~G., {Mathieu}, R.~D., {Vrba}, F.~J., {Mazeh}, T., \& {Henden}, A.
  2001, \aj, 121, 1003

\bibitem[{{Tilley} {et~al.}(2017){Tilley}, {Segura}, {Meadows}, {Hawley}, \&
  {Davenport}}]{Til17}
{Tilley}, M.~A., {Segura}, A., {Meadows}, V.~S., {Hawley}, S., \& {Davenport},
  J. 2017, ArXiv e-prints, arXiv:1711.08484

\bibitem[{{Tinker} {et~al.}(2002){Tinker}, {Pinsonneault}, \&
  {Terndrup}}]{Tin02}
{Tinker}, J., {Pinsonneault}, M., \& {Terndrup}, D. 2002, \apj, 564, 877

\bibitem[{{Tokunaga} {et~al.}(1990){Tokunaga}, {Toomey}, {Carr}, {Hall}, \&
  {Epps}}]{Tok90}
{Tokunaga}, A.~T., {Toomey}, D.~W., {Carr}, J., {Hall}, D.~N.~B., \& {Epps},
  H.~W. 1990, in \procspie, Vol. 1235, Instrumentation in Astronomy VII, ed.
  D.~L. {Crawford}, 131--143

\bibitem[{{Torres} {et~al.}(2010){Torres}, {Andersen}, \&
  {Gim{\'e}nez}}]{Tor10}
{Torres}, G., {Andersen}, J., \& {Gim{\'e}nez}, A. 2010, \aapr, 18, 67

\bibitem[{{Triaud} {et~al.}(2010){Triaud}, {Collier Cameron}, {Queloz},
  {Anderson}, {Gillon}, {Hebb}, {Hellier}, {Loeillet}, {Maxted}, {Mayor},
  {Pepe}, {Pollacco}, {S{\'e}gransan}, {Smalley}, {Udry}, {West}, \&
  {Wheatley}}]{Tri10}
{Triaud}, A.~H.~M.~J., {Collier Cameron}, A., {Queloz}, D., {et~al.} 2010,
  \aap, 524, A25

\bibitem[{{Uzdensky} {et~al.}(2002){Uzdensky}, {K{\"o}nigl}, \&
  {Litwin}}]{Uzd02}
{Uzdensky}, D.~A., {K{\"o}nigl}, A., \& {Litwin}, C. 2002, \apj, 565, 1191

\bibitem[{{van Eyken} {et~al.}(2011){van Eyken}, {Ciardi}, {Rebull},
  {Stauffer}, {Akeson}, {Beichman}, {Boden}, {von Braun}, {Gelino}, {Hoard},
  {Howell}, {Kane}, {Plavchan}, {Ram{\'{\i}}rez}, {Bloom}, {Cenko}, {Kasliwal},
  {Kulkarni}, {Law}, {Nugent}, {Ofek}, {Poznanski}, {Quimby}, {Grillmair},
  {Laher}, {Levitan}, {Mattingly}, \& {Surace}}]{van11}
{van Eyken}, J.~C., {Ciardi}, D.~R., {Rebull}, L.~M., {et~al.} 2011, \aj, 142,
  60

\bibitem[{{van Eyken} {et~al.}(2012){van Eyken}, {Ciardi}, {von Braun}, {Kane},
  {Plavchan}, {Bender}, {Brown}, {Crepp}, {Fulton}, {Howard}, {Howell},
  {Mahadevan}, {Marcy}, {Shporer}, {Szkody}, {Akeson}, {Beichman}, {Boden},
  {Gelino}, {Hoard}, {Ram{\'{\i}}rez}, {Rebull}, {Stauffer}, {Bloom}, {Cenko},
  {Kasliwal}, {Kulkarni}, {Law}, {Nugent}, {Ofek}, {Poznanski}, {Quimby},
  {Walters}, {Grillmair}, {Laher}, {Levitan}, {Sesar}, \& {Surace}}]{van12}
{van Eyken}, J.~C., {Ciardi}, D.~R., {von Braun}, K., {et~al.} 2012, \apj, 755,
  42

\bibitem[{{Vidotto} {et~al.}(2014){Vidotto}, {Gregory}, {Jardine}, {Donati},
  {Petit}, {Morin}, {Folsom}, {Bouvier}, {Cameron}, {Hussain}, {Marsden},
  {Waite}, {Fares}, {Jeffers}, \& {do Nascimento}}]{Vid14}
{Vidotto}, A.~A., {Gregory}, S.~G., {Jardine}, M., {et~al.} 2014, \mnras, 441,
  2361

\bibitem[{{Weber} \& {Davis}(1967)}]{Web67}
{Weber}, E.~J., \& {Davis}, Jr., L. 1967, \apj, 148, 217

\bibitem[{{Wurster} {et~al.}(2018){Wurster}, {Bate}, \& {Price}}]{Wur18}
{Wurster}, J., {Bate}, M.~R., \& {Price}, D.~J. 2018, \mnras, 481, 2450

\bibitem[{{Wyatt}(2008)}]{Wya08}
{Wyatt}, M.~C. 2008, \araa, 46, 339

\bibitem[{{Yadav} {et~al.}(2015{\natexlab{a}}){Yadav}, {Christensen}, {Morin},
  {Gastine}, {Reiners}, {Poppenhaeger}, \& {Wolk}}]{Yad15b}
{Yadav}, R.~K., {Christensen}, U.~R., {Morin}, J., {et~al.} 2015{\natexlab{a}},
  \apjl, 813, L31

\bibitem[{{Yadav} {et~al.}(2015{\natexlab{b}}){Yadav}, {Gastine},
  {Christensen}, \& {Reiners}}]{Yad15a}
{Yadav}, R.~K., {Gastine}, T., {Christensen}, U.~R., \& {Reiners}, A.
  2015{\natexlab{b}}, \aap, 573, A68

\bibitem[{{Yang} \& {Johns-Krull}(2011)}]{Yan11}
{Yang}, H., \& {Johns-Krull}, C.~M. 2011, \apj, 729, 83

\bibitem[{{Yang} {et~al.}(2005){Yang}, {Johns-Krull}, \& {Valenti}}]{Yan05}
{Yang}, H., {Johns-Krull}, C.~M., \& {Valenti}, J.~A. 2005, \apj, 635, 466

\bibitem[{{Yang} {et~al.}(2008){Yang}, {Johns-Krull}, \& {Valenti}}]{Yan08}
---. 2008, \aj, 136, 2286

\bibitem[{{Yu} {et~al.}(2017){Yu}, {Donati}, {H{\'e}brard}, {Moutou}, {Malo},
  {Grankin}, {Hussain}, {Collier Cameron}, {Vidotto}, {Baruteau}, {Alencar},
  {Bouvier}, {Petit}, {Takami}, {Herczeg}, {Gregory}, {Jardine}, {Morin},
  {M{\'e}nard}, \& {Matysse Collaboration}}]{Yu17}
{Yu}, L., {Donati}, J.-F., {H{\'e}brard}, E.~M., {et~al.} 2017, \mnras, 467,
  1342

\bibitem[{{Zanni} \& {Ferreira}(2013)}]{Zan13}
{Zanni}, C., \& {Ferreira}, J. 2013, \aap, 550, A99

\bibitem[{{Zwillinger} \& {Kokoska}(2000)}]{Zwi00}
{Zwillinger}, D., \& {Kokoska}, S. 2000, {CRC Standard Probability and
  Statistics Tables and Formulae. }

\end{thebibliography}



\end{document}